%% file: main.tex
\documentclass{article}
\usepackage[utf8]{inputenc}
\input{GrandMacros.tex}


\usepackage{amsthm}
\usepackage{xcolor}
\newtheorem{remark}{Remark}

\usepackage{graphicx,verbatim,array,multicol, subfigure, color, lscape, mathrsfs}
\usepackage{psfrag, amsmath, amsfonts, epsfig, fancybox, setspace, soul, amsthm, longtable, threeparttable, threeparttablex, makecell, xcolor, pdflscape, tabu, booktabs}
\usepackage[normalem]{ulem}

\usepackage[utf8]{inputenc}
\newcommand{\indep}{\perp \!\!\! \perp}

\def\sg{^{\text{gDiD}}}

\usepackage{setspace}
\title{Generalized difference-in-differences}
\author{Denis Agniel, Max Rubinstein, Jessie Coe, Maria DeYoreo}

\begin{document}

\newtheorem{assumption}{Assumption}
\newtheorem{prop}{Proposition}

\maketitle
\begin{abstract}
    We propose a new method for estimating causal effects in longitudinal/panel data settings that we call generalized difference-in-differences. Our approach unifies two alternative approaches in these settings: ignorability estimators (e.g., synthetic controls) and difference-in-differences (DiD) estimators. We propose a new identifying assumption -- a stable bias assumption -- which generalizes the conditional parallel trends assumption in DiD, leading to the proposed generalized DiD framework. This change gives generalized DiD estimators the flexibility of ignorability estimators while maintaining the robustness to unobserved confounding of DiD. We also show how ignorability and DiD estimators are special cases of generalized DiD. We then propose influence-function based estimators of the observed data functional, allowing the use of double/debiased machine learning for estimation. We also show how generalized DiD easily extends to include clustered treatment assignment and staggered adoption settings, and we discuss how the framework can facilitate estimation of other treatment effects beyond the average treatment effect on the treated. Finally, we provide simulations which show that generalized DiD outperforms ignorability and DiD estimators when their identifying assumptions are not met, while being competitive with these special cases when their identifying assumptions are met. 
\end{abstract}
\section{Introduction}

Researchers often use longitudinal or panel data to assess causal questions in public policy, economics, health services research, medicine, and other fields. These studies typically include units (e.g., employees, municipalities, patients) observed over time (e.g., months, quarters, years), where some of the units are exposed to an intervention or a treatment during the course of the study. For example, employees may participate in a training program, municipalities may enact a policy change, or patients may initiate a new treatment. 

To identify and estimate an average treatment effect with observational, non-experimental data, two main approaches exist. The first invokes an \textit{ignorability} assumption: that the treatment was effectively randomized within units sharing pre-treatment outcome and covariate values. The ignorability assumption allows for non-random treatment in that treatment can be related to all observed pre-treatment variables. The ignorability assumption often underpins various standard estimation approaches that effectively control for observed variables in estimation of a treatment effect, such as regression, propensity score weighting/matching, or synthetic controls. While this assumption allows for treated and untreated units to systematically differ, these differences are assumed only to occur among observed variables. In other words, we assume the researcher observes all confounding variables. 

The other main approach is a difference-in-differences (DiD) or event study design. Instead of relying on an ignorability assumption, DiD designs allow for some unobserved confounding between treated and untreated units. Specifically, DiD designs invoke a parallel trends assumption that allows for average outcomes between treated and untreated units to have different levels, as long as they would have the same changes over time absent treatment. When observed variables are thought to influence the evolution of outcomes, DiD designs hinge on a conditional parallel trends assumption. The conditional parallel trends assumption states that, within subgroups defined by levels of the observed variables, the average outcome of the treated units would have evolved in parallel to the average outcome of the untreated units had the treated units not received treatment. While the strain of literature related to estimation of treatment effects under an ignorability assumption has increasingly developed and adopted sophisticated machine learning techniques to account for high-dimensional control variables and non-linear functional forms, the DiD literature has largely relied on time-invariant controls and using generalized linear models for estimation. 

We propose a novel approach to causal identification and estimation of average treatment effect parameters from observational, longitudinal data that unifies these two strategies. We call our design ``generalized difference-in-differences'' (gDiD). The validity of gDiD relies on a \textit{stable bias assumption} which generalizes conditional parallel trends. While the conditional parallel trends assumption is generally stated as an equivalence between two groups, the stable bias assumption reformulates the assumption as an equivalence between two time periods. Conditional parallel trends is generally stated as: the conditional average difference over time in the outcomes of the treated units had they not been treated equals the conditional average difference over time in the outcomes of the untreated units, where the conditioning set is the same for both groups. The stable bias assumption states that the difference in conditional average outcomes of the treated units and untreated units after treatment had no one been treated equals the difference in conditional average outcomes of the treated units and untreated units before treatment. Reformulating the assumption as a comparison of time periods instead of groups clarifies how all pre-period information (for example, pre-period outcomes and other time-varying variables) can be accounted for in the conditioning. In particular, controlling for prior outcomes is straightforward. Like ignorability, the stable bias assumption allows for a rich dependence between the post-treatment outcomes and pre-period outcomes. Like parallel trends, the stable bias assumption allows for the presence of unmeasured confounding with stable effects over time. We further show that conditional and unconditional DiD estimators are special cases of gDiD, as are ignorability estimators under additional assumptions. 

The stable bias assumption states that the bias of an estimator in the post-treatment period can be estimated by examining the same estimator in the pre-treatment period. The identification result thus suggests debiasing an ignorability estimand in the post-treatment period by subtracting the corresponding estimand in the pre-treatment period.  In this paper, we focus on estimation of the average treatment effect on the treated (ATT). We propose an influence-function-based estimation strategy and provide the asymptotic distribution of the estimator. To implement our estimator, we make use of modern double/debiased machine learning techniques. In settings where the usual conditional parallel trends assumption holds, making use of these nonparametric estimation techniques makes standard DiD estimators robust to model mis-specification, and our simulation results suggest that this non-parametric approach remains competitive with DiD when parallel trends holds. Additionally, we provide simulation evidence that when conditional parallel trends holds but not all conditioning variables are observed to the researcher, a setting plausible in practice, our gDiD approach which includes prior outcomes in the conditioning set has lower bias than usual DiD estimators. 

Our contributions are thus threefold: first, we propose the gDiD design and show how the stable bias assumption provides a unified identification framework to think about both DiD estimators and ignorability estimators; second, we provide simulation evidence showing that the gDiD estimator performs well under various settings, including where ignorability and parallel trends hold; and third, our proposed estimation strategy, which can be implemented for standard DiD estimators as well as our gDiD estimator, employs the latest best practices in non-parametric machine learning techniques. Our accompanying R package implements these tools using SuperLearner to allow for flexible machine learning methods for estimation.

\subsection{Related work}

This work builds on recent advances in the difference-in-differences literature \cite{roth2023s}. Many recent estimators have generalized the classical parallel trends assumption \cite{callaway2021difference, sant2020doubly, borusyak2021revisiting}. These conditional parallel trends assumptions allow the relationship between trends in the counterfactual outcome under no intervention to have a complex dependence on covariates. We further generalize these assumptions to depend on pre-period covariates -- in such a way that they're no longer ``parallel trends" assumptions, but related ``stable bias" assumptions. This allows for more complex relationships between post-period (counterfactual) outcomes and pre-period outcomes. Furthermore, this additional conditioning on pre-period outcomes can minimize the effect of missing a covariate that would otherwise be required to ensure conditional parallel trends (see Section \ref{sims} for simulation evidence in this setting).

We also build on the application of synthetic control techniques \cite{abadie2021using, ben2021augmented} to the setting where many units may be treated. While the assumptions of synthetic control estimators are often written in terms of an assumed model (e.g., a linear factor model), they can alternatively be justified by an ignorability assumption conditional on the pre-treatment outcomes. Augmented synthetic controls \cite{ben2021augmented} can furthermore be seen as a special case of an augmented inverse probability weighted estimator \cite{bang2005doubly, chernozhukov2018double} where the pre-treatment outcomes are treated as confounders. Generalized difference-in-differences relaxes the ignorability assumption these estimators make, while still allowing practitioners to take advantage of the full suite of ignorability tools -- i.e., synthetic control estimation or double machine learning. 

Finally, this work addresses the concerns raised by Daw and Hatfield \cite{daw2018matching} that controlling for pre-period outcomes can lead to bias in DiD settings because of regression to the mean. They specifically consider a caliper matching strategy, where units in the treated sample that do not have a suitable match in the control sample are discarded. Related work by Ham and Miratrix \cite{ham2022benefits} suggests that matching on pre-period outcomes can improve on traditional DiD and provide conditions under a simple model where this improvement can be ensured. The proposed gDiD framework explicitly conditions on pre-period outcomes to achieve causal identification and establishes conditions under which this is justified. We argue that the required stable bias assumption is much more flexible than parallel trends (and thus likely in general more plausible), but we also show how this additional flexibility requires a stricter positivity (or overlap) assumption. The setting of Daw and Hatfield is an example of a failure of our positivity assumption, which may be investigated using standard causal inference tools.

The rest of the paper proceeds as follows. In Section \ref{main-section}, we outline the basic problem setup and the assumptions we require to identify the target causal effect, including the stable bias assumption. We compare our identifying assumptions to the corresponding assumptions (ignorability, parallel trends, and conditional parallel trends) for alternative estimators. We also give the influence function for the identifying functional of the ATT under our assumptions, use it to propose an estimator, and derive the limiting normal distribution of the estimator. In Section \ref{extensions-section}, we outline extensions of our approach to staggered adoption and clustered treatment assignment settings, and we discuss extensions to other types of treatment effects beyond the ATT. In Section \ref{sims}, we give extensive simulation results for our approach in comparison to competing DiD and ignorability estimators. We apply gDiD to estimate the causal effect of health system affiliation on health care quality among Medicare Fee-for-service beneficiaries in Section \ref{section:real-data}. Finally, we give concluding remarks in Section \ref{discussion}.

\section{Generalized difference-in-differences}\label{main-section}

\subsection{Preliminaries and basic assumptions}\label{preliminaries-section}

We observe $n$ units and assume $\bO = (\bO_i)_{i = 1, ..., n} = (Y_{i1}, A_i, \bX_i, \bYbar_{i0})_{i = 1, ..., n}$ are a set of iid random variables. For simplicity, we consider a single post-treatment time period $t=1$ and a binary treatment $A_i = \{0,1\}$ such that no units are treated before period 1 and some units are treated in period 1. $Y_{it}$ denotes the outcomes at time $t$, which we group into pre-treatment outcomes up to time 0, $\bYbar_{i0} = (Y_{it})_{t \leq 0}$, and post-treatment outcomes, $Y_{i1}$. The $\bX_i$ are a set of time-invariant covariates.\footnote{While our notation implies that the covariates are time-invariant, $\bX_i$ may also include vectors of time-varying covariates measured prior to treatment, similar to $\bYbar_{i0}$.} Let $\bW_{it} = (\bX_i, \bYbar_{it})$ for $t \leq 0$ be the set of covariates and pre-treatment outcomes at time $t$.

We make use of potential outcomes notation \cite{rubin2005causal} and define the potential outcome $Y_{i1}(a)$ as the outcome at $t=1$ that one would observe for unit $i$ if, possibly contrary to fact, the treatment took value $A_i = a$ for  $a = \{0, 1\}$. 

Our primary causal estimand of interest is the average treatment effect on the treated (ATT) at time-period $t = 1$:

\begin{align*}
    \tau = E\{Y_{i1}(1) - Y_{i1}(0) | A_i = 1\}
\end{align*}
This is a conventional estimand in policy evaluation settings where we observe longitudinal or panel data.\footnote{We later extend this to staggered adoption settings in the vein of \cite{callaway2021difference} in Section \ref{staggered} as well as to clustered treatment assignment in Section \ref{clustered-design}. We will also later consider other estimands like the average treatment effect (not limited to the treated group) in Section \ref{other-trt-effects}.}

To identify this estimand from the observed data, we require the following standard assumptions:
\begin{assumption}[Consistency]\label{asmp:consistency}
    $Y_{i1} = A_iY_{i1}(1) + (1-A_i)Y_{i1}(0)$. 
\end{assumption}
\begin{assumption}[No anticipation]\label{asmp:anticipation}
    $Y_{it}(a) = Y_{it}, a = 0, 1; t < 1.$
\end{assumption}
\begin{assumption}[Positivity]\label{asmp:positivity}
    $P\left\{ \pi(\bW_{i0}) < 1 - \epsilon\right\} = 1$ for the propensity score $\pi(\bw) = P(A_i = 1 | \bW_{i0} = \bw)$ and some $\epsilon > 0$.
\end{assumption}
Assumption \ref{asmp:consistency} allows us to connect the observed data with counterfactual outcomes and states that we observe the counterfactual outcome under treatment $A_i = a$ when $A_i$ takes value $a$. Assumption \ref{asmp:anticipation} ensures no anticipatory effects of treatment prior to the intervention and states that counterfactuals in the pre-treatment period are observed for treatment and control. Finally, Assumption \ref{asmp:positivity} ensures that there is sufficient overlap in the treated and control groups and states that there is at least some probability of finding units like the treated units among the controls.
%
%

\subsection{Stable bias assumption and identification}\label{identification-section}

The crucial identifying assumption for our generalized difference-in-differences approach is the stable bias assumption:
\begin{assumption}[Stable bias]\label{stable-bias}
    \begin{align*}E\left[E\left\{Y_{i1}(0) | \bW_{i0}, A_i = 1\right\} - E\left\{Y_{i1} | \bW_{i0}, A_i = 0 \right\} | A_i = 1\right] = E\left[E\left\{Y_{i0} | \bW_{i-1}, A_i = 1\right\} - E\left\{Y_{i0} | \bW_{i-1}, A_i = 0\right\} | A_i = 1\right].\end{align*}
\end{assumption}
This assumption states that for counterfactual outcomes under no treatment, the difference in conditional average outcomes of the treated units and untreated units at $t=1$ (after treatment) equals the corresponding difference at $t=0$ (before treatment).

Under Assumptions \ref{asmp:consistency}, \ref{asmp:anticipation}, \ref{asmp:positivity}, and \ref{stable-bias}, $\tau$ may be identified in the observed data via the following expression:
\begin{align}
\begin{aligned}\label{tau-gdid}
    \tau^{\text{gDiD}} =&E(Y_{i1} | A_i = 1) - E\left\{ E(Y_{i1} | \bW_{i0}, A_i = 0) | A_i = 1 \right\} - \\
    &\qquad \left\{E(Y_{i0} | A_i = 1) - E\left\{ E(Y_{i0} | \bW_{i-1}, A_i = 0) | A_i = 1 \right\}\right\}
    \end{aligned}
\end{align}

This observed data functional has a difference-in-differences structure. The first differences is the difference in the mean post-treatment observed outcomes among the treated units and the mean post-treatment outcomes among the control units, if the control units had a distribution of $\bW_{i0}$ similar to the treated units. The second difference is the difference in the mean observed outcomes at $t=0$ among the treated units and the control units, if the control units had a distribution of $\bW_{i-1}$ similar to the treated units. The first difference is the identifying functional of the ATT under the assumption that $\bW_{i0}$ is sufficient to control for confounding (Assumption \ref{ignorability-assmp} below) -- see $\tau^{\text{ign.}}$ given in \eqref{tau-ign}. The second component is equivalent to a similar ignorability estimand using only pre-treatment data. The form of the expression - a difference in these two estimands - takes the form of a DiD estimator but one which uses pre-period outcomes in the conditioning set. 

\subsection{Assumptions and identification of previous approaches}\label{prev-approaches}

Versions of Assumptions 1-3 have been required by essentially all estimators in this setting. Differences arise between estimators in the additional assumption required for identifying $\tau$ from the observed data. We briefly review the two standard assumptions (justifying ignorability and DiD estimators, respectively) and discuss how our stable bias assumption relates to previous approaches. 

Ignorability estimators of the ATT require the following assumption
\begin{assumption}[Conditional ignorability]\label{ignorability-assmp}
    $Y_{i1}(0) \indep A_i | \bW_{i0}$.
\end{assumption}
\noindent The assumption states that conditional on everything known prior to treatment, the treatment is as if randomized with respect to the counterfactual outcome under no treatment at $t = 1$. This assumption is alternatively known as \textit{selection on observables} or \textit{no unmeasured confounding}. Under Assumptions \ref{asmp:consistency}, \ref{asmp:anticipation}, \ref{asmp:positivity}, and \ref{ignorability-assmp}, $\tau$ is identified from the observed data by
\begin{align}\begin{aligned}\label{tau-ign}
    \tau^{\text{ign.}} &= E\left\{E(Y_{i1} | \bW_{i0}, A_i = 1) -  E(Y_{i1} | \bW_{i0}, A_i = 0) | A_i = 1 \right\} \\
 &= E(Y_{i1} |  A_i = 1) -  E\left\{E(Y_{i1} | \bW_{i0}, A_i = 0) | A_i = 1 \right\} .
\end{aligned}
\end{align}
Compare $ \tau^{\text{ign.}}$ to $ \tau^{\text{gDiD}}$. Note that $ \tau^{\text{gDiD}}$ can be considered as a difference in ignorability estimators - one applied to the post-treatment period and one applied to the pre-treatment period. While the conditional ignorability assumption (Assumption \ref{ignorability-assmp}) assumes that the bias of $ \tau^{\text{ign.}}$ is zero, the proposed stable bias assumption (Assumption \ref{stable-bias}) assumes that the bias of $\tau^{\text{ign.}}$ may be non-zero and may be estimated in the pre-period data. Note that ignorability implies stable bias as long as there is no pre-period treatment effect: $E(Y_{i0} | A_i = 1) - E\left\{ E(Y_{i0} | \bW_{i-1}, A_i = 0) | A_i = 1 \right\} = 0$, which in turn is equivalent to there being no unmeasured pre-treatment confounding.

All DiD estimators require some form of the parallel trends assumption. The typlical conditional parallel trends assumption states that the mean counterfactual trend in the outcome absent treatment in the treated group is equal to the observed trend in the control group within strata defined by time-invariant covariates:
\begin{assumption}[Conditional parallel trends]\label{parallel-trends-cond}
$E\left\{Y_{i1}(0) - Y_{i0}(0) | \bX_i, A_i = 1\right\} = E\left\{Y_{i1}(0) - Y_{i0}(0) | \bX_i, A_i = 0\right\}.$    
\end{assumption}
Consider a toy example of health status and gender. The traditional conditional parallel trends assumption (Assumption \ref{parallel-trends-cond}) assumes that the health status of treated women if they hadn't been treated would evolve similarly to the health status of untreated women. Our stable bias assumption (Assumption \ref{stable-bias}) allows the evolution of health status to depend on initial health status and accounts for effects of past outcomes on present outcomes. In parallel trends language, our stable bias assumption assumes that in a counterfactual setting without treatment, the health status of treated women with low pre-treatment ($t < 0$) health would evolve in parallel to the health status of untreated women with low pre-treatment health, while accounting for potential associations between $Y_{i0}$ and $Y_{i1}$.  

Further, because the usual identification requires only conditioning on time-invariant $\bX_i$, the version of the positivity assumption required by classic DiD estimators is slightly weaker:
\begin{assumption}[Time-invariant covariate positivity]\label{asmp:positivity-ti}
    $P\left\{ \pi_X(\bX_{i}) < 1 - \epsilon\right\} = 1$ for the propensity score $\pi_X(\bx) = P(A_i = 1 | \bX_i = \bx)$ and some $\epsilon > 0$.
\end{assumption}
Under Assumptions \ref{asmp:consistency}, \ref{asmp:anticipation}, \ref{parallel-trends-cond}, and \ref{asmp:positivity-ti}, $\tau$ is identified from the observed data by
\begin{align*}
 \tau^{\text{cDiD}} &= E\left\{E(Y_{i1} | \bX_i, A_i = 1) - E(Y_{i0} \mid \bX_i, A_i = 1) - (E(Y_{i1} | \bX_i, A_i = 0)  - E(Y_{i0} \mid \bX_i, A_i = 0)) | A_i = 1 \right\} \\
 &= E(Y_{i1} | A_i = 1) -  E\left\{E(Y_{i1} | \bX_i, A_i = 0)| A_i = 1 \right\} - E(Y_{i0} | A_i = 1) + E\left\{E(Y_{i0} \mid \bX_i, A_i = 0) | A_i = 1 \right\}.
\end{align*}

Consider a further comparison of these three approaches to identifying causal effects in this setting. All of these estimators of the ATT are imputation estimators of the form:
\begin{align*}
    \tau = E\{Y_{i1}(1) - Y_{i1}(0) | A_i = 1\} = E(Y_{i1} | A_i = 1) - E\{\eta(\bW_{i0}) | A_i = 1\}
\end{align*}
where by assumption the counterfactual $E\{Y_{i1}(0) | A_i = 1\}$ equals $E\{\eta(\bW_{i0}) | A_i = 1\}$, which is identified from the observed data. 

Assumption \ref{ignorability-assmp} (ignorability) leads to 
    $\eta^{\text{ign.}}(\bW_{i0}) = E(Y_{i1} | \bW_{i0}, A_i = 0)$,
while under parallel trends (Assumption \ref{parallel-trends-cond}), the imputation function looks like $\eta^{\text{cDiD}}(\bW_{i0}) = Y_{i0} + \Delta(\bX_i)$
for $\Delta(\bX_i) = E(Y_{i1} - Y_{i0} | \bX_i, A_i = 0)$, the conditional trend in the untreated group. Under stable bias (Assumption \ref{stable-bias}), the imputation function is
    $\eta^{\text{gDiD}}(\bW_{i0}) = Y_{i0} + \Delta^*(\bW_{i0})$
for $\Delta^*(\bW_{i0}) = E(Y_{i1} | \bW_{i0}, A_i = 0) - E(Y_{i0} | \bW_{i-1}, A_i = 0) = \eta^{\text{ign.}}(\bW_{i0}) - E(Y_{i0} | \bW_{i-1}, A_i = 0)$.

Comparing these imputation functions reveals that the complexity of the ignorability imputation function is essentially unbounded -- it can depend nonlinearly on $Y_{i0}$ and may encode interactions between $Y_{i0}, \bYbar_{i-1},$ and $\bX_i$ -- while the DiD imputation function is in comparison quite simple. It depends strictly linearly on $Y_{i0}$ and can encode no interactions between $Y_{i0}$ and $\bX_i$ and ignores information in previous pre-period outcomes completely. While the generalized DiD imputation function includes a term that is linear in $Y_{i0}$, it also includes $\Delta^*(\bW_{i0})$, a term that can capture nonlinearity and interactions in the relationship between $\bYbar_{i0}$ and $Y_{i1}$ and is of similar complexity to $\eta^{\text{ign.}}$.

\begin{remark}
    We argue that the gDiD positivity and stable bias assumptions (Assumptions \ref{asmp:positivity} and \ref{stable-bias}) in comparison to the positivity and parallel trends assumptions of DiD (Assumptions \ref{asmp:positivity-ti} and \ref{parallel-trends-cond}) represent a kind of bias-variance trade-off. The differences between these sets of assumptions lie in the conditioning set. If pre-period outcomes are ignored and $\bW_{it}$ is taken to be $\bX_i$, then Assumption \ref{asmp:positivity} is equivalent to Assumption \ref{asmp:positivity-ti}, and Assumption \ref{stable-bias} is equivalent to Assumption \ref{parallel-trends-cond}. 

    We have further just argued that by using Assumption \ref{stable-bias}, gDiD is able to use a much more complex imputation function for the counterfactual $E\{Y_{i1}(0) | A_i = 1\}$. The price of this additional complexity in the imputation function -- which should in general lead to lower bias -- is to make the positivity assumption stronger. If there is no or only weak overlap in the pre-period outcomes, Assumption \ref{asmp:positivity} may be less likely to hold than Assumption \ref{asmp:positivity-ti}. When Assumption \ref{asmp:positivity} is nearly violated, this can lead to inflated variance in gDiD (as well as in related ignorability estimators). We demonstrate this numerically in Section \ref{sims}. Note also that there is a similar trade-off required to move from the unconditional parallel trends assumption to Assumption \ref{parallel-trends-cond}: the bias reduction due to conditioning on $\bX_i$ comes at the cost of the assumption of overlap on $\bX_i$. There is, as they say, no free lunch.
\end{remark}

\subsection{Estimation and inference}\label{if-and-estimation}

Having identified $\tau$ in terms of the observed data distribution, estimation of $\tau$ amounts to efficient estimation of the expectations 
    $E\{\mu_{1}(\bW_{i0}) | A_i = 1\} = E\left\{ E(Y_{i1} | \bW_{i0}, A_i = 0) | A_i = 1 \right\}$
    and 
        $E\{\mu_{0}(\bW_{i-1}) | A_i = 1\} = E\left\{ E(Y_{i0} | \bW_{i-1}, A_i = 0) | A_i = 1 \right\}$.
We propose using an influence-function based approach. The influence functions for these functionals \cite{chernozhukov2017double} can be expressed as
\begin{align*}
    & E(A_i)\inv\left[\frac{(1-A_i)\pi(\bW_{i0})Y_{i1} + \{A_i - \pi(\bW_{i0})\}\mu_1(\bW_{i0})}{1-\pi(\bW_{i0})} - E\{\mu_1(\bW_{i0}) | A_i = 1\}\right] \text{ and }\\
    & E(A_i)\inv\left[\frac{(1-A_i)\pi_0(\bW_{i-1})Y_{i0} + \{A_i - \pi_0(\bW_{i-1})\}\mu_0(\bW_{i-1})}{1-\pi_0(\bW_{i-1})} - E\{\mu_0(\bW_{i-1}) | A_i = 1\} \right]
\end{align*}
where $\pi_0(\bW_{i-1}) = P(A_i = 1 | \bX_i = \bx, \bYbar_{i-1} = \bybar_{-1}).$ Because the efficient influence function for a sum is the sum of the influence functions, we have therefore that the influence function for \eqref{tau-gdid} can be written as

\begin{align}
\begin{aligned}\label{phi}
    \phi(\bO_i) &= E(A_i)\inv\left[Y_{i1}A_i - \frac{(1-A_i)\pi(\bW_{i0})Y_{i1} + \{A_i - \pi(\bW_{i0})\}\mu_1(\bW_{i0})}{1-\pi(\bW_{i0})} - \right.\\
    &\qquad\qquad \left.Y_{i0}A_i +  \frac{(1-A_i)\pi_0(\bW_{i-1})Y_{i0} + \{A_i - \pi_0(\bW_{i-1})\}\mu_0(\bW_{i-1})}{1-\pi_0(\bW_{i-1})}\right] - \tau\sg.
\end{aligned}
\end{align}

Thus, a doubly-robust estimator for $\tau$ may be constructed as
\begin{align}
\begin{aligned}
    \tauhat &= n_1\inv\sumin \left[Y_{i1}A_i - \frac{(1-A_i)\pihat(\bW_{i0})Y_{i1} + \{A_i - \pihat(\bW_{i0})\}\muhat_1(\bW_{i0})}{1-\pihat(\bW_{i0})} - \right.\\
    &\qquad\qquad \left.Y_{i0}A_i +  \frac{(1-A_i)\pihat_0(\bW_{i-1})Y_{i0} + \{A_i - \pihat_0(\bW_{i-1})\}\muhat_0(\bW_{i-1})}{1-\pihat_0(\bW_{i-1})}\right]
\end{aligned}\label{tauhat}
\end{align}
where $\pihat, \pihat_0, \muhat_1, \muhat_0$ are estimates of the nuisance functions $\pi, \pi_0, \mu_1, \mu_0$ and $n_1 = \sumin A_i$.
There are two predominant strategies for estimating the nuisance functions. When the dataset is small, parametric models may be specified for the nuisance functions, 
%
as in, for example, \cite{callaway2021difference} and \cite{sant2020doubly}. However, parametric models are difficult to specify and often incur bias due to model misspecification, so when the dataset is not small we propose to use non-parametric machine learning methods to estimate the nuisance functions ($\pi, \pi_0, \mu_1, \mu_0$) in combination with cross-fitting \cite{chernozhukov2018double}. Cross-fitting separates the estimation of the nuisance functions from their evaluation in \eqref{tauhat}, allowing only weak assumptions on the complexity of the underlying nuisance functions to achieve root-$n$ consistent and asymptotically normal estimators of the causal effect \cite{kennedy2022semiparametric}. Intuitively, cross-fitting prevents a kind of overfitting that might otherwise occur when using non-parametric estimation methods.

While cross-fitting has been described extensively elsewhere (see, e.g., \cite{kennedy2022semiparametric}), we briefly review its implementation. Specifically, we first randomly splitting the data into $K$ disjoint subsets (indexed from $k = 1, ..., K$) with $N_k$ observations, respectively. Next, for each subset, one estimates the nuisance functions on all data except the $k$-th subset and constructs the estimator $\hat{\tau}_k$ on the the $k$th subset. Finally, the cross-fit estimator is constructed as the sample-size weighted average of each of the $K$ estimators, i.e. $\sum_{k=1}^K\frac{N_k}{N}\hat{\tau}_k$.  

Under minimal assumptions on the nuisance function estimates, $\tauhat$ will have an asymptotic normal distribution. This means that we may do standard inference and construct confidence intervals for $\tau$, even when using machine learning methods for estimation. Specifically, we employ the following assumption on the rates of convergence of the nuisance function estimates:
\begin{assumption}[Nuisance function convergence]
\begin{align}\label{asymptotic-convergence-assump}
    &E\left[ \left\{\muhat_1(\bW_{i0}) - \mu_1(\bW_{i0} \right\}^2 \right]^{1/2} \times E\left[ \left\{\pihat(\bW_{i0}) - \pi(\bW_{i0} \right\}^2 \right]^{1/2} = o_p(\nnhalf)\\
    &E\left[ \left\{\muhat_0(\bW_{i-1}) - \mu_0(\bW_{i-1} \right\}^2 \right]^{1/2} \times E\left[ \left\{\pihat_0(\bW_{i-1}) - \pi_0(\bW_{i-1} \right\}^2 \right]^{1/2} = o_p(\nnhalf)
\end{align}
\end{assumption}
The cross-fitting approach we have proposed in Section \ref{if-and-estimation} along with most off-the-shelf machine learning estimators is generally sufficient to ensure Assumption \ref{asymptotic-convergence-assump}, as has been noted in other contexts previously \cite{chernozhukov2018double}, such as for regression trees/random forests\cite{wager2015adaptive} or deep learning\cite{farrell2021deep}. This allows us to state the limiting distribution of the proposed estimator
\begin{prop}\label{asymptotic-distribution}
Under Assumptions \ref{asmp:consistency}, \ref{asmp:anticipation}, \ref{asmp:positivity}, \ref{stable-bias}, \ref{asymptotic-convergence-assump}, and additional regularity conditions in Appendix \ref{app:prop-1}, 
\begin{align}\label{normal-dist}
    n^{-1/2}\sigma\inv(\tauhat - \tau) \longrightarrow N(0, 1), \qquad \sigma^2 = E\{\phi(\bO_i)^2\}.
\end{align}
\end{prop}
The proof of this proposition is ensured by previous results for the ATT \cite{chernozhukov2018double, kennedy2022semiparametric}, as our proposed estimator is the difference between two ATTs. We give details in Appendix \ref{app:prop-1}. And, as with previous results, Proposition 1 continues to hold if $\sigma$ is replaced by $\sigmahat$ defined by
\begin{align}\label{sigmahat}
    \sigmahat^2 &= n_1\inv\sumin \left(\left[Y_{i1}A_i - \frac{(1-A_i)\pihat(\bW_{i0})Y_i + \{A_i - \pihat(\bW_{i0})\}\muhat_1(\bW_{i0})}{1-\pihat(\bW_{i0})} - \right.\right.\\
    &\qquad\qquad \left.\left.Y_{i0}A_i +  \frac{(1-A_i)\pihat_0(\bW_{i-1})Y_i + \{A_i - \pihat_0(\bW_{i-1})\}\muhat_0(\bW_{i-1})}{1-\pihat_0(\bW_{i-1})}\right] - \tauhat\right)^2
\end{align}
The immediate consequence of Proposition \ref{asymptotic-distribution} is that a (1-$\alpha$)\% confidence interval can therefore be constructed as $\tauhat \pm z_{1-\alpha/2}\sigmahat$ where $z_{\gamma}$ is the $\gamma$th quantile of the standard normal distribution. We review an alternative multiplier bootstrap variance estimator in Appendix \ref{app:bootstrap}. 

\begin{remark}
    Assumption \ref{asymptotic-convergence-assump} will not be ensured when parametric models are used to estimate nuisance functions without sample splitting unless both models are correctly specified. In Appendix \ref{app:parametric} we give a general form for the asymptotic variance under no assumptions about correct model specification. Furthermore, we also note that it would be straightforward to compute design-based variances in the style of \cite{rambachan2020design}, which may be of interest in settings where units cannot be thought of as arising from a super-population of units.
\end{remark}

\begin{remark}
    The use of non-parametric estimation methods in combination with influence-function based estimators has frequently been discouraged in the DiD literature. For example, \cite{sant2020doubly} propose using generalized linear models in combination with their influence-function based DiD estimator, arguing that the ``curse of dimensionality'' makes non-parametric methods less useful in practice. While we agree with this caution in a small-sample setting, non-parametric estimation methods can yield asymptotically normal and root-n consistent estimates under relatively weak conditions, even in a high-dimensional setting. For example, as noted by \cite{kennedy2022semiparametric}, if we assume that all true model propensity-score and regression models lie within some $s$-sparse model class, lasso-based estimates of the nuisance parameters can achieve root-n consistent and asymptotically normal estimates provided $s = o(\sqrt{n} / \log d)$ and that we're able to achieve $O_p(\sqrt{\frac{s \log d}{n}})$ nuisance estimation rates in the $L_2$ norm. By contrast, when using parametric models in combination with influence-function based parametric estimators, all models are likely misspecified in practice leading to inconsistent estimators. Influence-based estimators using parametric models for the nuisance parameters may be relatively less useful than non-parametric models. We therefore recommend using modern non-parametric and machine methods in combination with these estimators when the sample size is not small. Our accompanying R package implements our proposed method using SuperLearner, an ensemble of machine learning methods; however, our package can also accommodate generalized linear models for estimation if desired. 
\end{remark}

\section{Extensions}\label{extensions-section}

\subsection{Staggered adoption and group-time treatment effects} \label{staggered}

In settings where units may adopt or be exposed to the treatment at different times (the so-called staggered adoption setting), the basic framework of generalized DiD may be extended to estimate ATTs that capture potentially heterogeneous treatment effects over time. We first establish some additional notation.

Let $\bA_i = (A_{it})_{t=1, ..., T}$ be a vector of treatment indicators over time, where now time is indexed from $t = 1$ to $t = T$. Let $\Asc_i$ be the time of adoption $\Asc_i = \min_{A_{it} = 1} t$ if any $A_{it} = 1$ and is $\infty$ otherwise, and let $\Psc_i = \Asc_i - 1$ be the last pre-period time point. We also redefine the counterfactuals so that they potentially depend on the entire treatment history prior to the current time, $Y_{it}(\babar_t)$ for $\babar_t = (a_s)_{s = 1, .., t}$, and let $\Asc_{\babar_t}$ be the implied time of adoption for a given treatment history $\babar_t$: $\Asc_{\babar_t} = \min_{a_s = 1} s$. Define $\bAbar_{it} = (A_{is})_{s =1 ,..., t}$, and let $\Psc_{\babar_t} = \Asc_{\babar_t} - 1$.

Then we may define the group-time treatment effect
\begin{align*}
    \tau_{\babar_t} = E\{Y_{it}(\babar_t) - Y_{it}(\bzero_t) | \bAbar_{it} = \babar_t\}.
\end{align*}
Notice that $\tau_{\babar_t}$ actually generalizes the estimands considered in \cite{callaway2021difference} as it places no restrictions on the treatment history. That is, we do not require that once a unit is treated it remains treated. 

The assumptions required to identify $\tau_{\babar_t}$ are as follows. First, we update the consistency assumption for the new counterfactuals,

\begin{assumption}[Consistency, staggered adoption setting]\label{asmp:consistency-sa}
    $Y_{it}(\babar_t) = Y_{it} \text{ when } \bAbar_t = \babar_t.$
\end{assumption}
\begin{assumption}[No anticipation, staggered adoption setting]\label{asmp:anticipation-sa}
    $Y_{is}(\babar_t) = Y_{is},  \text{ when } s < \Asc(\babar_t)$.
\end{assumption}
\begin{assumption}[Positivity, staggered adoption setting]\label{asmp:positivity-sa}
    $P\left\{ \pi_{\babar_t}(\bW_{i\Psc_{\babar_t}}) < 1 - \epsilon\right\} = 1$, for $\pi_{\babar_t}(\bw) = P(\bAbar_{it} = \babar_t | \bW_{i\Psc_{\babar_t}} = \bw)$.
\end{assumption}
\begin{assumption}[Stable bias, staggered adoption setting]\label{stable-bias-sa}
    \begin{align*}
    \begin{aligned}
    &E\left[E\left\{Y_{it}(\bzero_t) | \bW_{i\Psc_{\babar_t}}, \bAbar_{it} = \babar_t\right\} - E\left\{Y_{it} | \bW_{i\Psc_{\babar_t}}, \bAbar_{it} = \bzero_t \right\} | \bAbar_{it} = \babar_t\right] = \\
    &\qquad E\left[E\left\{Y_{i0} | \bW_{i\Psc_{\babar_t}-1}, \bAbar_{it} = \babar_t\right\} - E\left\{Y_{i0} | \bW_{i\Psc_{\babar_t}-1}, \bAbar_{it} = \bzero_t\right\} | \bAbar_{it} = \babar_t\right].
    \end{aligned}
\end{align*}
\end{assumption}
These assumptions directly generalize Assumptions \ref{asmp:consistency}, \ref{asmp:anticipation}, \ref{asmp:positivity}, and \ref{stable-bias} and serve the same purposes. Under Assumptions \ref{asmp:consistency-sa}, \ref{asmp:anticipation-sa}, \ref{asmp:positivity-sa}, and \ref{stable-bias-sa}, the group-time ATT $\tau_{\babar_t}$ may be estimated from the observed data via the following functional
\begin{align}
\begin{aligned}
    E\{Y_{it}(\babar_t) - Y_{it}(\bzero_t) | \bAbar_{it} = \babar_t\} &= E(Y_{it} | \bAbar_{it} = \babar_t) - E\left\{ E(Y_{it} | \bW_{i0}, A_i = 0) | \bAbar_{it} = \babar_t \right\} - \\
    &\qquad E(Y_{i0} | \bAbar_{it} = \babar_t) + E\left\{ E(Y_{i0} | \bW_{i-1}, A_i = 0) | \bAbar_{it} = \babar_t \right\}
    \end{aligned}\label{grouptime-tau-identified}
\end{align}

The influence function and estimation for \eqref{grouptime-tau-identified} follow directly from Section \ref{if-and-estimation}, and the asymptotics follow precisely the same arguments as in that section. 

We can also aggregate these functionals over time, following \cite{callaway2021difference}. Let $\Xi_t \subset \{0, 1\}^t$ be a set of treatment histories such that 
\begin{align*}
    \Xi_t = \left\{ \abar_t : P(\Abar_{it} = \babar_t) > 0 \text{ and Assumptions \ref{asmp:consistency-sa}, \ref{asmp:anticipation-sa}, \ref{asmp:positivity-sa}, and \ref{stable-bias-sa} hold.}\right\}
\end{align*}
Further let $\bomega_t = \{\omega_{\babar_t}\}_{\babar_t \in \Xi_t}$ be a set of weights defined over the elements $\Xi_t$. Then we can define aggregated treatment effects as
\begin{align}\label{aggregated-effects}
    \tau_{\bomega_t} = \sum_{\babar_t \in \Xi_t} \omega_{\babar_t} \tau_{\babar_t}
\end{align}
Examples of weights leading to interpretable treatment effects include, the average effect of the intervention in year $t$ among units that are treated at year $t$, given by $\omega_{\babar_t} = I\{a_t = 1\}/\sum_{\babar_t \in \Xi_t}I\{a_t = 1\}$. Another example is the effect of the intervention in time $t$ among units that adopted treatment at time $s$, given by $\omega_{\babar_t} = I\{\Asc(\babar_t) = s\}/\sum_{\babar_t \in \Xi_t}I\{\Asc(\babar_t) = s\}$. A final example is the effect of the intervention at time $t$ among units that adopted treatment at time $s$ and were also treated at time $t$, given by $\omega_{\babar_t} = I\{\Asc(\babar_t) = s, a_t = 1\}/\sum_{\babar_t \in \Xi_t}I\{\Asc(\babar_t) = s, a_t = 1\}$.

\begin{remark}
    Just as in Section \ref{prev-approaches}, if the pre-period outcomes are omitted from the covariate vector, then Assumption \ref{stable-bias-sa} reduces to an assumption similar to the parallel trends assumption of \cite{callaway2021difference}. However, we are using a generalized version of their estimand allowing identification of treatment effects when units may switch between treatment and control group unrestrictedly. 
\end{remark}

\subsection{Clustered treatment assignment} \label{clustered-design}

In many settings we observe clusters $j = 1, ..., C$; and units $i = 1, ..., n_j$ within each group, where the $j$th of $C$ clusters has $n_j$ observations indexed by $i$. Treatment $A_j$ is then assigned to all units within each cluster. We assume that the data in the $j$th cluster $\bO_j = \left\{Y_{ji1}(0), Y_{ji1}(1), \bX_{ji}, A_j, \bYbar_{ji0}\right\}$ is independent of the data in the $k$th cluster ($k \neq j$), $\bO_k = \left\{Y_{ki1}(0), Y_{ki1}(1), \bX_{ki}, A_k, \bYbar_{ki0}\right\}$, and, for simplicity, we assume that unit-cluster assignment is fixed (i.e. units cannot change clusters). We define the ATT at time-period $t = 1$ to be:
\begin{align}\label{clustered-tau}
    \tau = E\{Y_{ji1}(1) - Y_{ji1}(0) | A_j = 1\}.
\end{align}

The outcome in this setting may depend on characteristics of the individual observation captured in $\bWbar_{ji0} = (\bX_{ji}, \bYbar_{ji0}),$ and it may also depend on characteristics of the other members of the cluster $\{\bWbar_{jk0} : k \neq i\}$. Define a set of cluster-level summary variables $\bWtilde_{j0}, \bWtilde_{j-1}$ -- e.g., $\bWtilde_{jt} = n_j\inv\sum_{i=1}^{n_j}\bW_{jit}$ -- and collect both the individual- and cluster-level information into the vector $\bV_{jit} = (\bW_{jit}, \bWtilde_{jt})$. Let $N_1 = \sum_{j=1}^C n_jA_j$ be the total number of treated observations. Then we can write an influence-function based estimator for \eqref{clustered-tau} as
\begin{align}
    \begin{aligned}
    \tauhat &= N_1\inv\sum_{j=1}^C \sum_{i=1}^{n_j} \left[Y_{ji1}A_j - \frac{(1-A_j)\phat(\bWtilde_{j0})Y_{ji1} + \{A_j - \phat(\bWtilde_{j0})\}\mhat_1(\bV_{ji0})}{1-\phat(\bWtilde_{j0})} - \right.\\
    &\qquad\qquad \left.Y_{ji0}A_j +  \frac{(1-A_j)\phat_0(\bWtilde_{j-1})Y_{ji0} + \{A_j - \phat_0(\bWtilde_{j-1})\}\mhat_0(\bV_{ji-1})}{1-\phat_0(\bWtilde_{j-1})}\right]
    &= C\inv\sum_{j=1}^C \phitilde(\bO_j)
\end{aligned}\label{clustered-tauhat}
\end{align}
where $\phat(\bv)$ is an estimate of $ P(A_j = 1 | \bV_{ji0} = \bv)$, $\phat_0(\bv)$ is an estimate of $P(A_j = 1 | \bV_{ji-1} = \bv)$, and $\phitilde(\bO_j)$ is an uncentered version of the influence function
\begin{align*}
    \phi(\bO_j) &= \frac{C}{N_1}\sum_{i=1}^{n_j} \left[Y_{ji1}A_j - \frac{(1-A_j)p(\bV_{ji0})Y_{ji1} + \{A_j - p(\bV_{ji0})\}m_1(\bV_{ji0})}{1-p(\bV_{ji0})} - \right.\\
    &\qquad\qquad \left.Y_{ji0}A_j +  \frac{(1-A_j)p_0(\bV_{ji-1})Y_{ji0} + \{A_j - p_0(\bV_{ji-1})\}m_0(\bV_{ji-1})}{1-p_0(\bV_{ji-1})} - \tau\right],
\end{align*}

\begin{prop}
    Under Assumptions \ref{asmp:consistency-cl}, \ref{asmp:anticipation-cl}, \ref{asmp:positivity-cl}, \ref{stable-bias-cl}, \ref{asmp:dgp-reg-cl}, and \ref{asmp:nuisance-reg-cl} in Appendix \ref{app:clustered}, 
    \begin{align*}
        C^{-1/2}(\tauhat - \tau) \longrightarrow N(0, \zeta^2)
    \end{align*}
    for $\zeta^2 = E\{\phi(\bO_j)^2\}$.

\end{prop}
Proof given in Appendix \ref{app:clustered}. This proposition states that the estimator under clustered treatment assignment has a similar asymptotic distribution to \eqref{normal-dist}, the distribution for the estimator in the iid case. A plug-in version of $\zeta^2$ may be computed just as in \eqref{sigmahat}, but we also show in Appendix \ref{app:bootstrap} that a multiplier bootstrap estimator may equivalently be used.

\subsection{Other treatment effects} \label{other-trt-effects}

Traditional DiD and the parallel trends assumption have been engineered specifically to estimate the ATT, but our alternative formulation of the stable bias assumption opens the door to estimating many other, equally or more relevant effects. For example, the Centers of Medicare \& Medicaid Services may implement a voluntary intervention that is adopted by only a subset of eligible insurance plans participating in the Medicare Advantage program. The ATT estimates the expected effect of the program if other similar plans also adopted the intervention. However, policymakers may also be interested in the effect if all plans were to implement the intervention, which would be captured by the so-called \textit{average treatment effect} (ATE):
\begin{align*}
    \tau_{\text{ATE}} = E\{Y_{i1}(1) - Y_{i1}(0)\}.
\end{align*}
Similarly, in the example in Section \ref{section:real-data}, we examine the effect of health system affilliation on quality of care delivered by physician organizations (POs) to Medicare beneficiaries. POs that affiliate with health systems tend to be larger and tend to have fewer racial and ethnic minority beneficiaries. Policymakers may not be interested particularly in this subset of POs but may prefer to estimate effects of health system affiliation in all POs.

The assumptions needed to identify the ATE in the context of unmeasured confounding and the tools needed to estimate it are nearly identical as to the ATT, which we explored in detail above. The positivity assumption (Assumption \ref{asmp:positivity} must be updated to be a little stricter, requiring positive probability of being either treated or untreated for all covariate strata, and the stable bias assumption must also be updated:
\begin{assumption}[Positivity, ATE]
    $P\{\epsilon < \pi(\bW_{i0}) < 1 - \epsilon\} = 1$
\end{assumption}
\begin{assumption}[Stable bias, ATE]
\begin{align*}
    E\left\{E(Y_{i1} | A = 1, \bW_{i0}) -  E(Y_{i1} | A = 0, \bW_{i0})\right\} - \tau_{\text{ATE}} &= 
    E\left\{E(Y_{i0} | A = 1, \bW_{i-1}) -  E(Y_{i0} | A = 0, \bW_{i-1})\right\}.
    \end{align*}
\end{assumption}
Under these two assumptions, as well as Assumptions \ref{asmp:consistency} and \ref{asmp:anticipation}, we may estimate the ATE in a manner analogous to \eqref{tauhat}.

\begin{remark}
    While we argue that alternative estimands -- like the ATE -- may have more policy relevance in some cases, stable bias assumptions for these alternatives may be harder to justify. Intuitively, pre-treatment data arguably provides information about $Y_{i1}(0)$; however, it is harder to argue that these same data provide information about $Y_{i1}(1)$. For causal estimands such as the ATE that effectively require imputing $Y_{i1}(1)$ for untreated units, it requires further study to establish that pre-treatment biases with respect to $Y_{it}(0)$ (for $t < 1$) can inform post-treatment biases with respect to averages of these unobserved counterfactuals. Nevertheless, these sorts of stable bias assumptions may be reasonable in some settings, and in these cases we can follow the logic above to identify and estimate these causal quantities.
\end{remark}


\section{Simulations}\label{sims}
We conduct two complementary sets of simulation studies. The first closely follows the setup from \cite{sant2020doubly} and compares the performance of the proposed gDiD estimator to state-of-the-art DiD estimation methods. However, the space of possible data-generating mechanisms in the panel data setting is much broader than considered in \cite{sant2020doubly}. We therefore conduct a second set of simulation studies that allows a more thorough exploration of the performance of different estimators under a range of assumptions about the data-generating processes. For all simulations, we fit all models on five-hundred random datasets.

We consider the following estimators:
\begin{itemize}
    \item Three implementations of gDiD: one where we control for only one pre-period outcome (gDiD-1) -- $\bW_{it} = (\bX_i, Y_{it})$ -- one where we control for two pre-period outcomes (gDiD-2) -- $\bW_{it} = (\bX_i, Y_{it}, Y_{it-1})$ -- and one where we only use time-invariant covariates (gDiD-0) -- $\bW_{it} = \bX_i$. The gDiD-0 estimator is similar to the estimator in \cite{callaway2021difference} and \cite{sant2020doubly} but using machine learning models instead of parametric models. 
    \item Two ignorability estimators, controlling for one (Ign-1) and two (Ign-2) pre-period outcomes, using augmented inverse probability weighting.
    \item Augmented synthetic controls \cite{ben2021augmented, augsynth-package} controlling for three pre-period outcomes. This is essentially an additional (regression-based) ignorability estimator using regularized linear models \cite{bruns2023augmented}.
    \item A standard DiD estimator.
    \item The doubly-robust conditional DiD estimator given by \cite{sant2020doubly}. 
\end{itemize}
For the Ign and gDiD estimators, we use cross-fitting and a weighted combination of GLMs, Random Forests, and GAMs, where the weights on each model are learned using SuperLearner \cite{van2007super, polley2019package}. For all estimators except for augmented synthetic controls, we conduct inference using the empirical variance of the estimated influence functions and standard normal quantiles. We do not conduct inference for augmented synthetic controls because there are not enough pre-treatment time-periods to use the default conformal inference based approach \cite{chernozhukov2021exact} implemented in the \texttt{augsynth} package. Below, we report bias, root mean squared error (RMSE), average 95\% confidence interval length (CIL), and 95\% confidence interval coverage (Coverage). 

Before getting into details about the simulations, we briefly give motivation and high-level results. In the first simulation, we modify the simulations in \cite{sant2020doubly} such that a variable required for conditional parallel trends is not available. Here we show that, while DiD estimators incur bias due to the omitted variable, gDiD and ignorability estimators are able to perform well because pre-period outcomes are able to proxy for the omitted variable. We further show how positivity violations can degrade the performance of ignorability estimators and how, in this setting, gDiD suffers less from positivity issues than ignorability estimators.

In the second simulation setting, we start from a setting where parallel trends holds but ignorability does not and show that gDiD is competitive with DiD (ignorability estimators perform poorly). We then go on to vary the simulation parameters away from parallel trends in various ways and show that (generally speaking) gDiD outperforms all other estimators (often by a factor of two or more in terms of RMSE) no matter how we vary the simulation.

\subsection{First data-generating process}
\subsubsection{Setup}
We first conduct a simulation study that closely follows \cite{sant2020doubly}. However, we make two key modifications to their specification: first, we change the data-generating process to change the levels of overlap in pre-treatment outcomes; second, we consider a setting where conditional parallel trends does not hold given the observed covariates. To be precise, we consider a generic four-dimensional covariate $\bW = (W_1, W_2, W_3, W_4)$ and define regression and propensity score functions:
$f_{reg}(\bW) = 205 + 27.4 W_1 + 13.7 (W_2 + W_3 + W_4)$, 
$f_{ps}(\bW) = 0.75 (-W_1 + 0.5W_2 - 0.5W_3 - 0.25W_4)$.
%
We take $n = 1,000$ independent draws of a four-dimensional covariate vector $\bX$ from $MVN(0, I_4)$, where $I_4$ is a 4x4 identity matrix. We then take nonlinear transformations of $\bX$, which we call $\tilde{\bZ}$, and then standardize $\tilde{\bZ}$ to obtain $\bZ$. Specifically, define $\tilde{Z}_1 = \exp(0.5X_1)$, $\tilde{Z}_2 = 10 + X_2 / (1 + \exp(X_1))$, $\tilde{Z}_3 = (0.6 + X_1X_3/25)^3$, and $\tilde{Z}_4 = (20 + X_2 + X_4)^2$. For each covariate $j = 1, ..., 4$, we define $Z_j = (\tilde{Z}_j - \mathbb{E}[\tilde{Z}_j]) / \sqrt{Var(\tilde{Z}_j)}$. For $t = 1, ..., 4$, we build on \cite{kang2007demystifying} and \cite{sant2020doubly} and consider the following data-generating process:
$Y_t(1) = Y_t(0) = 0.1tf_{reg}(\bX) + v(A, \bX; \zeta) + \epsilon_t$, 
$A = 1(\text{expit}(f_{ps}(\bX)) \ge U)$, 
%
where $\epsilon_t$ is an independent standard normal random variable, $U$ is an independent random uniform variable, and $v(A, \bX; \zeta)$ is an independent normal random variable with mean $f_{reg}(\bX)(1 + \zeta A)$ and variance one. We change the parameter $\zeta$ to explore different levels of overlap in the pre-treatment outcomes. When there is less overlap in the pre-treatment outcomes, Assumption \ref{asmp:positivity} becomes (nearly) violated. 

We consider two sets of observed data $\bO = (\bO_i)_{i=1, ...n}$, which we call the \textit{linear} and \textit{non-linear} cases. In the linear case, we consider data where the true covariates $X_2, X_3, X_4$ are available for analysis $\bO_i = (Y_{i-2}, ..., Y_{i1i}, A_i, X_{i2}, ..., X_{i4})$. In the non-linear case, only the transformed covariates $Z_2, Z_3, Z_4$ are available but not the $X$s: $\bO_i = (Y_{i-2}, ..., Y_{i1}, A_i, X_{i1}, ..., X_{i3})$. In both cases the observed data do not include the first covariate $X_{i1}$ or $Z_{i1}$. In contrast to the simulation in \cite{sant2020doubly}, this simulation includes time-varying unobserved heterogeneity due to omission of $X_1$ and $Z_1$. As a result, conditional parallel trends does not hold in our simulations, though it would if we observed all four covariates.\footnote{Notice that $X_1$ has the strongest confounding effect on the outcome as it has the strongest effect on both the treatment assignment and outcome models.} We allow $\zeta \in \{0, 0.1, 0.3, 0.5\}$, where the higher we set $\zeta$ the less overlap in the pre-treatment outcome levels between the treatment and control groups. When $\zeta = 0$, both ignorability and conditional parallel trends would hold were we to observe all covariates. We choose this specification because we believe that in practice parallel trends is unlikely to hold exactly, as we are not likely to measure all relevant confounders.

\subsubsection{Results}


Table~\ref{tab:sim1a} presents the results in both the linear and nonlinear settings when $\zeta = 0$. The results illustrate that the estimators that control for pre-period outcomes (gDiD-1, gDiD-2, Ign-1, Ign-2, Augsynth) have bias close to zero in both settings. This suggests that controlling for pre-period outcomes is sufficient to control for confounding, while the estimators that rely strictly on parallel trends suffer from bias. While the gDiD-1 and gDiD-2 estimators have low bias, their RMSEs are somewhat higher than the ignorability-based estimation approaches (Ign-1, Ign-2, and AugSynth). The additional bias correction from gDiD appears unnecessary in this simulation, and the estimation of the bias therefore induces additional variability and no bias. The worst performing estimators in terms of bias are the DiD, cDiD, and gDiD-0 estimators. Interestingly, all three of these estimators perform comparably, possibly a function of the relatively small confounding effects of the included covariates. Overall, these results illustrate how gDiD and ignorability estimators that take advantage of pre-period outcomes can overcome omitted variables in some settings and how gDiD is competitive with ignorability estimators when ignorability holds.

\begin{table}\caption{Simulation study one results: $\zeta = 0$}
\centering
\begin{tabular}{l>{}l>{}lll>{}l>{}lll}
\toprule
\multicolumn{1}{c}{ } & \multicolumn{4}{c}{Linear} & \multicolumn{4}{c}{Nonlinear} \\
\cmidrule(l{3pt}r{3pt}){2-5} \cmidrule(l{3pt}r{3pt}){6-9}
Estimator & Bias & RMSE & CIL & Coverage & Bias & RMSE & CIL & Coverage\\
\midrule
DID & -0.21 & 0.312 & 0.966 & 88.2 & -0.224 & 0.344 & 0.965 & 84\\
cDID & -0.194 & 0.284 & 0.788 & 83.7 & -0.237 & 0.349 & 0.899 & 80.3\\
gDID-0 & -0.199 & 0.289 & 0.794 & 82.8 & -0.235 & 0.333 & 0.87 & 80.2\\
Ign-1 & -0.003 & 0.101 & 0.38 & 93.8 & -0.006 & 0.099 & 0.388 & 96\\
Ign-2 & \textbf{0.002} & \textbf{0.088} & 0.334 & 93.2 & -0.006 & \textbf{0.085} & 0.335 & 94.2\\
AugSynth & 0.006 & 0.185 & - & - & -0.002 & 0.149 & - & -\\
gDID-1 & -0.004 & 0.172 & 0.657 & 94 & 0.005 & 0.175 & 0.666 & 95.2\\
gDID-2 & 0.003 & 0.134 & 0.508 & 94.5 & -0.002 & 0.133 & 0.51 & 94.8\\
\bottomrule
\end{tabular}\label{tab:sim1a}
\end{table}

Table~\ref{tab:sim1b} presents results when setting $\zeta = 0.1$. Note that in this setting none of the key identifying assumptions (ignorability, conditional parallel trends, or stable bias) hold. Changing $\zeta$ also causes a decreased level of overlap between the pre-treatment outcomes in the treated and control groups and causes a near violation of Assumption \ref{asmp:positivity}. These changes do not effect the DiD estimators but have two implications for the performance of the gDiD and ignorability estimators. First, because stable bias and ignorability do not hold, these estimators incur much more bias than when $\zeta = 0$. Second, because Assumption \ref{asmp:positivity} is closer to being violated, the variances and associated confidence intervals get larger. These changes lead to unacceptably large biases in the ignorability estimators, but the gDiD estimators remain competitive with the DiD and cDiD estimators, with gDiD-1 performing the best in both the linear and non-linear settings. In  Appendix \ref{app:sim-1} we display results setting $\zeta = 0.3, 0.5$. The results are consistent with this trend, with worsening performance among the gDiD and Ign estimators, and the performance of the cDiD and DiD estimators remaining consistent.

\begin{table}\caption{Simulation study one results: $\zeta = 0.1$}
\centering
\begin{tabular}{l>{}l>{}lll>{}l>{}lll}
\toprule
\multicolumn{1}{c}{ } & \multicolumn{4}{c}{Linear} & \multicolumn{4}{c}{Nonlinear} \\
\cmidrule(l{3pt}r{3pt}){2-5} \cmidrule(l{3pt}r{3pt}){6-9}
Estimator & Bias & RMSE & CIL & Coverage & Bias & RMSE & CIL & Coverage\\
\midrule
DID & -0.219 & 0.333 & 0.966 & 85.8 & -0.213 & 0.322 & 0.964 & 85.8\\
cDID & -0.194 & 0.278 & 0.788 & 86 & -0.229 & 0.322 & 0.887 & 83.5\\
gDID-0 & -0.197 & 0.285 & 0.794 & 85 & -0.225 & 0.317 & 0.87 & 82.8\\
Ign-1 & -1.549 & 1.553 & 0.489 & 0 & -1.553 & 1.556 & 0.487 & 0\\
Ign-2 & -2.381 & 2.386 & 0.656 & 0 & -2.392 & 2.396 & 0.652 & 0\\
AugSynth & -3.218 & 3.252 & - & - & -3.252 & 3.28 & - & -\\
gDID-1 & \textbf{0.128} & \textbf{0.227} & 0.707 & 89.8 & \textbf{0.13} & \textbf{0.226} & 0.698 & 87.8\\
gDID-2 & 0.221 & 0.317 & 0.788 & 78.7 & 0.215 & 0.324 & 0.771 & 78.8\\
\bottomrule
\end{tabular}\label{tab:sim1b}
\end{table}

\subsection{Second data-generating process}
\subsubsection{Setup}
The previous simulations make somewhat restrictive assumptions on the data-generating processes. For example, it is unlikely that coefficients on the unobserved covariates change linearly with time. Additionally, prior period outcomes may directly enter the outcome model. However, the above specification becomes complex when incorporating these nuances due to the presence of covariates. We instead consider a setting where the data-generating process in each time period is only a function of one unobserved covariate and one prior-period outcome, with no other observed covariates entering the outcome model. Specifically, we generate data according to the following general model:
$Y_t(1) = Y_t(0) = \beta_t f(\theta) + \gamma_t g(Y_{t-1}(0)) + (t - 1) + \epsilon_t, t = -2, -1, 0, 1$, 
$A = 1(\text{expit}(f(\theta)) > U)$
%
where $\theta$, $Y_{-2}$, and $\epsilon_t$ are all independent standard normal random variables.

We consider several parameterizations of $(\beta_t, \gamma_t, f, g, \alpha)$. First, consider the possible specification of $\gamma_t$, which govern the association between the current outcome and previous outcomes:
\begin{enumerate}
    \item No prior-period effects: $\gamma_t = \gamma = 0$
    \item Constant prior-period effects: $\gamma_t = \gamma \neq 0$
    \item Linear effect growth: $\gamma_t - \gamma_{t-1} = c > 0$
    \item Random effect changes: $\gamma_t - \gamma_{t-1} = c_t \neq 0$
\end{enumerate}
We have the same set of possible specifications for $\beta_t$, except that we do not consider $\beta_t = 0$. Additionally, we consider linear and additive specifications (a third nonlinear specification is given in Appendix \ref{app:sim-2}). Linear corresponds to $f, g$ being the identity function, and additive corresponds to $f(x) = \text{sign}(x)\exp(\lvert \sin(x) \rvert)(\lvert x\rvert > 1) + (0.9x + x^3)(\lvert x\rvert <= 1)$ and $g(x) = (x - x^2)(\lvert x\rvert <= 1.5) + 2\text{sign}(x)\log(\lvert x\rvert)(\lvert x\rvert > 1.5)$ and $g(x) = (x - x^2)(\lvert x\rvert <= 1.5) + 2\text{sign}(x)\log(\lvert x\rvert)(\lvert x\rvert > 1.5)$. 

\begin{remark}
    These simulation settings were chosen to explore a wide range of potential data-generating processes and not to simulate under the required assumptions of any of the methods. Thus it is not expected that bias will be 0 for any of the estimators necessarily in most settings.
\end{remark}

\subsubsection{Results}


First consider the results for the linear specification, given in Table \ref{sim-table-2}. when previous outcomes do not directly effect the current outcome ($\gamma_t = \gamma = 0$) and the effect of $\theta$ is constant over time $\beta_t = \beta$ (the conventional DiD setting), traditional DiD performs the best. Even in this setting the gDiD methods are competitive with DiD. Augmented synthetic controls and the ignorability estimators have RMSEs that are 2-4 times higher then gDiD. Outside of this setting, the gDiD-1 and gDiD-2 estimators outperform all other estimators by at least a factor of about 2 in terms of RMSE, with two exceptions. 

Results for the additive setting are similar and are given in Table \ref{sim-table-3}. Results for the nonlinear setting are also generally similar, though there are more instances of violations of Assumption \ref{asmp:positivity}, which degrades the performance of gDiD and ignorability estimators dramatically. See Appendix \ref{app:sim-2}.

\begin{landscape}
    
\begin{longtable}{l>{}l>{}lll>{}l>{}lll}
\toprule
\multicolumn{1}{c}{ } & \multicolumn{4}{c}{$\beta_t = \beta$} & \multicolumn{4}{c}{$\beta_t - \beta_{t-1} = d$} \\
\cmidrule(l{3pt}r{3pt}){2-5} \cmidrule(l{3pt}r{3pt}){6-9}
Estimator & Bias & RMSE & CIL & Coverage & Bias & RMSE & CIL & Coverage\\
\midrule
\endfirsthead
\multicolumn{9}{@{}l}{\textit{(continued)}}\\
\toprule
Estimator & Bias & RMSE & CIL & Coverage & Bias & RMSE & CIL & Coverage\\
\midrule
\endhead

\endfoot
\bottomrule
\endlastfoot
\addlinespace[0.3em]
\multicolumn{9}{l}{\textbf{$\gamma_t = \gamma = 0$}}\\
\hspace{1em}DID & -0.001 & \textbf{0.087} & 0.351 & 94.8 & 0.422 & 0.432 & 0.368 & 0.8\\
\hspace{1em}gDID-0 & \textbf{-0.001} & 0.105 & 0.401 & 94 & 0.244 & 0.266 & 0.418 & 38\\
\hspace{1em}Ign-1 & 0.456 & 0.464 & 0.349 & 0 & 0.593 & 0.603 & 0.445 & 0\\
\hspace{1em}Ign-2 & 0.315 & 0.327 & 0.344 & 5.8 & 0.46 & 0.473 & 0.425 & 1.4\\
\hspace{1em}AugSynth & 0.24 & 0.255 & 0.963 & - & 0.425 & 0.439 & 1.701 & -\\
\hspace{1em}gDID-1 & 0.011 & 0.13 & 0.496 & 92.8 & \textbf{-0.066} & \textbf{0.168} & 0.616 & 93.4\\
\hspace{1em}gDID-2 & 0.005 & 0.126 & 0.486 & 94.6 & -0.132 & 0.197 & 0.577 & 83.6\\
\addlinespace[0.3em]
\multicolumn{9}{l}{\textbf{$\gamma_t = \gamma = c$}}\\
\hspace{1em}DID & 0.833 & 0.837 & 0.336 & 0 & 1.625 & 1.63 & 0.512 & 0\\
\hspace{1em}gDID-0 & 0.39 & 0.4 & 0.347 & 1.2 & 1.01 & 1.018 & 0.489 & 0\\
\hspace{1em}Ign-1 & 0.287 & 0.3 & 0.345 & 10.8 & 0.561 & 0.572 & 0.446 & 0.6\\
\hspace{1em}Ign-2 & 0.281 & 0.296 & 0.344 & 12.2 & 0.455 & 0.468 & 0.427 & 2.2\\
\hspace{1em}AugSynth & 0.273 & 0.285 & 1.091 & - & 0.434 & 0.448 & 1.736 & -\\
\hspace{1em}gDID-1 & -0.096 & 0.158 & 0.467 & 84.4 & -0.202 & 0.242 & 0.507 & 64.4\\
\hspace{1em}gDID-2 & \textbf{-0.081} & \textbf{0.157} & 0.483 & 89 & \textbf{-0.19} & \textbf{0.24} & 0.58 & 76.4\\
\addlinespace[0.3em]
\multicolumn{9}{l}{\textbf{$\gamma_t - \gamma_{t-1} = c$}}\\
\hspace{1em}DID & 1.38 & 1.385 & 0.468 & 0 & 2.276 & 2.282 & 0.669 & 0\\
\hspace{1em}gDID-0 & 0.414 & 0.425 & 0.378 & 1.2 & 1.029 & 1.039 & 0.544 & 0\\
\hspace{1em}Ign-1 & 0.253 & 0.268 & 0.342 & 18.8 & 0.47 & 0.483 & 0.434 & 2.4\\
\hspace{1em}Ign-2 & 0.255 & 0.27 & 0.343 & 17 & 0.437 & 0.452 & 0.43 & 3.4\\
\hspace{1em}AugSynth & 0.252 & 0.265 & 1.008 & - & 0.44 & 0.452 & 1.759 & -\\
\hspace{1em}gDID-1 & -0.072 & 0.141 & 0.474 & 89.6 & \textbf{-0.181} & \textbf{0.23} & 0.535 & 72.4\\
\hspace{1em}gDID-2 & \textbf{-0.069} & \textbf{0.14} & 0.476 & 89.8 & -0.192 & 0.246 & 0.575 & 74.2\\*
\caption{Simulation results showing the good performance of the proposed gDiD estimator in comparison to ignorability and DiD estimators. Performance is measured in terms of bias, root mean squared error (RMSE), confidence interval length (CIL), and 95\% confidence interval coverage (``coverage"). Columns indicate whether the effect of the unobserved parameter $\theta$ is constant over time ($\beta_t = \beta$) or whether it changes over time ($\beta_t - \beta_{t-1} = d$). Results are also separated by the effect of the previous timepoint's outcome: no effect of prior-period outcome ($\gamma_t = \gamma = 0$), constant prior-period-outcome effect ($\gamma_t = \gamma = c$), and changing prior-period-outcome effect ($\gamma_t - \gamma_{t-1} = c)$. Results are aggregated over 500 simulations of the linear specification.}
\end{longtable}\label{sim-table-2}

\begin{longtable}{l>{}l>{}lll>{}l>{}lll}
\caption{Simulation results showing the good performance of the proposed gDiD estimator in comparison to ignorability and DiD estimators. Performance is measured in terms of bias, root mean squared error (RMSE), confidence interval length (CIL), and 95\% confidence interval coverage (``coverage"). Columns indicate whether the effect of the unobserved parameter $\theta$ is constant over time ($\beta_t = \beta$) or whether it changes over time ($\beta_t - \beta_{t-1} = d$). Results are also separated by the effect of the previous timepoint's outcome: no effect of prior-period outcome ($\gamma_t = \gamma = 0$), constant prior-period-outcome effect ($\gamma_t = \gamma = c$), and changing prior-period-outcome effect ($\gamma_t - \gamma_{t-1} = c)$. Results are aggregated over 500 simulations of the additive (non-linear) specification.}\\
\toprule
\multicolumn{1}{c}{ } & \multicolumn{4}{c}{$\beta_t = \beta$} & \multicolumn{4}{c}{$\beta_t - \beta_{t-1} = d$} \\
\cmidrule(l{3pt}r{3pt}){2-5} \cmidrule(l{3pt}r{3pt}){6-9}
Estimator & Bias & RMSE & CIL & Coverage & Bias & RMSE & CIL & Coverage\\
\midrule
\endfirsthead
\multicolumn{9}{@{}l}{\textit{(continued)}}\\
\toprule
Estimator & Bias & RMSE & CIL & Coverage & Bias & RMSE & CIL & Coverage\\
\midrule
\endhead

\endfoot
\bottomrule
\endlastfoot
\addlinespace[0.3em]
\multicolumn{9}{l}{\textbf{$\gamma_t = \gamma = 0$}}\\
\hspace{1em}DID & 0.019 & \textbf{0.093} & 0.357 & 94.6 & 0.387 & 0.398 & 0.375 & 1\\
\hspace{1em}gDID-0 & \textbf{0.017} & 0.101 & 0.337 & 90.8 & 0.188 & 0.212 & 0.344 & 45\\
\hspace{1em}Ign-1 & 0.401 & 0.409 & 0.328 & 0 & 0.493 & 0.503 & 0.406 & 0\\
\hspace{1em}Ign-2 & 0.285 & 0.297 & 0.321 & 6.4 & 0.385 & 0.398 & 0.388 & 3.4\\
\hspace{1em}AugSynth & 0.228 & 0.241 & 0.911 & - & 0.413 & 0.424 & 1.651 & -\\
\hspace{1em}gDID-1 & 0.023 & 0.126 & 0.489 & 94.8 & \textbf{-0.083} & \textbf{0.169} & 0.593 & 93.8\\
\hspace{1em}gDID-2 & 0.023 & 0.121 & 0.467 & 94.8 & -0.133 & 0.193 & 0.545 & 81.6\\
\addlinespace[0.3em]
\multicolumn{9}{l}{\textbf{$\gamma_t = \gamma = c$}}\\
\hspace{1em}DID & 0.103 & 0.148 & 0.427 & 84 & 0.623 & 0.634 & 0.474 & 0\\
\hspace{1em}gDID-0 & 0.218 & 0.242 & 0.376 & 38.4 & 0.573 & 0.584 & 0.414 & 0.4\\
\hspace{1em}Ign-1 & 0.368 & 0.378 & 0.372 & 1.8 & 0.631 & 0.641 & 0.474 & 0.2\\
\hspace{1em}Ign-2 & 0.337 & 0.35 & 0.366 & 4.8 & 0.572 & 0.585 & 0.464 & 0.4\\
\hspace{1em}AugSynth & 0.122 & 0.169 & 0.558 & - & 0.437 & 0.456 & 1.75 & -\\
\hspace{1em}gDID-1 & \textbf{-0.034} & \textbf{0.134} & 0.528 & 96 & -0.098 & 0.178 & 0.589 & 89\\
\hspace{1em}gDID-2 & -0.055 & 0.138 & 0.497 & 93.2 & \textbf{-0.077} & \textbf{0.168} & 0.577 & 91\\
\addlinespace[0.3em]
\multicolumn{9}{l}{\textbf{$\gamma_t - \gamma_{t-1} = c$}}\\
\hspace{1em}DID & 0.283 & 0.307 & 0.451 & 32.6 & 0.771 & 0.781 & 0.51 & 0\\
\hspace{1em}gDID-0 & 0.192 & 0.227 & 0.405 & 52.6 & 0.5 & 0.517 & 0.441 & 2\\
\hspace{1em}Ign-1 & 0.333 & 0.348 & 0.391 & 7.6 & 0.527 & 0.539 & 0.488 & 0.8\\
\hspace{1em}Ign-2 & 0.303 & 0.319 & 0.385 & 12 & 0.498 & 0.511 & 0.477 & 0.6\\
\hspace{1em}AugSynth & 0.05 & 0.145 & 0.464 & - & 0.381 & 0.41 & 1.525 & -\\
\hspace{1em}gDID-1 & \textbf{-0.02} & 0.145 & 0.539 & 93.8 & -0.089 & 0.177 & 0.591 & 91.4\\
\hspace{1em}gDID-2 & -0.033 & \textbf{0.139} & 0.505 & 94.2 & \textbf{-0.072} & \textbf{0.174} & 0.569 & 90.4\\*
\end{longtable}\label{sim-table-3}
\end{landscape}

\section{Estimating the effect of health system affiliation on quality of care}\label{section:real-data}

Health systems, commonly known to be a group of affiliated health care providers including hospitals and physician organizations, have comprised an increasing share of the health care market in recent years\cite{furukawa2020consolidation}. Because health systems are still relatively new, the effect of health system affiliation on the quality of care delivered by physician organizations is still uncertain.

We use Medicare Fee-for-service claims data from 2013-2018 to study the effect of health system affiliation on four measures of the quality of care received by Medicare beneficiaries. These four quality measures come from the Health Effectiveness Data and Information Set and are approved by the National Quality Forum. Health system affiliation was determined by examining the Medicare Provider Enrollment, Chain, and Ownership System and Internal Revenue Service Form 990 database. Beneficiaries were attributed to physician organizations (POs, identified by Tax Identification Numbers) at which they received the plurality of their primary care visits during the calendar year. 

To estimate the effect of affiliating in year $t, t = 2015, 2016, 2017, 2018$ on outcomes in that same year, we let $A_j =1$ for POs who were first affiliated in year $t$ and $A_j = 0$ for POs who never affiliated during our study period. This is a clustered data setting (Section \ref{clustered-design}) because beneficiaries are clustered withiin POs and POs were the units affiliating with health systems. Beneficiaries were included in the analysis of year $t$ if they were eligible for the quality measure and were attributed to the same TIN in each of the following years: $t - 2, t-1, t$. For each year $t$, we estimated \eqref{clustered-tau}, taking the outcome in year $t$ to be $Y_{ji1}$, and the outcome in year $t -1$ to be $Y_{ji0}$ for beneficiary $i$ in PO $j$.

We adjusted for the following beneficiary-level characteristics: HCC risk score (a measure of health status), dual eligibility for Medicaid and Medicare (a measure of low income), social deprivation index (a measure of socioeconomic status in the beneficiary's community), age, race, place of residence (rural, urban, or small town), and an indicator of Medicare eligibility due to disability prior to aging into eligibility. In addition to these beneficiary-level covariates, we adjusted for PO-level aggregates of the same characteristics, as well as the following PO-level characteristics: number of attributed beneficiaries, number of physicians, and an indicator of including both primary care and specialists in the PO. To manage computation time, we subsampled within each affiliating and non-affiliating TIN, up to 1,000 beneficiaries in each affiliating TIN and up to 50 beneficiaries in each non-affiliating TIN. We used the multiplier bootstrap outlined in  Appendix \ref{app:bootstrap} for all standard errors. We also estimated aggregated effects, similar to \eqref{aggregated-effects}, combining the estimated effects across years of affiliation to arrive at a single affiliation effect.

\begin{figure}
\begin{center}
    \includegraphics[scale=0.6]{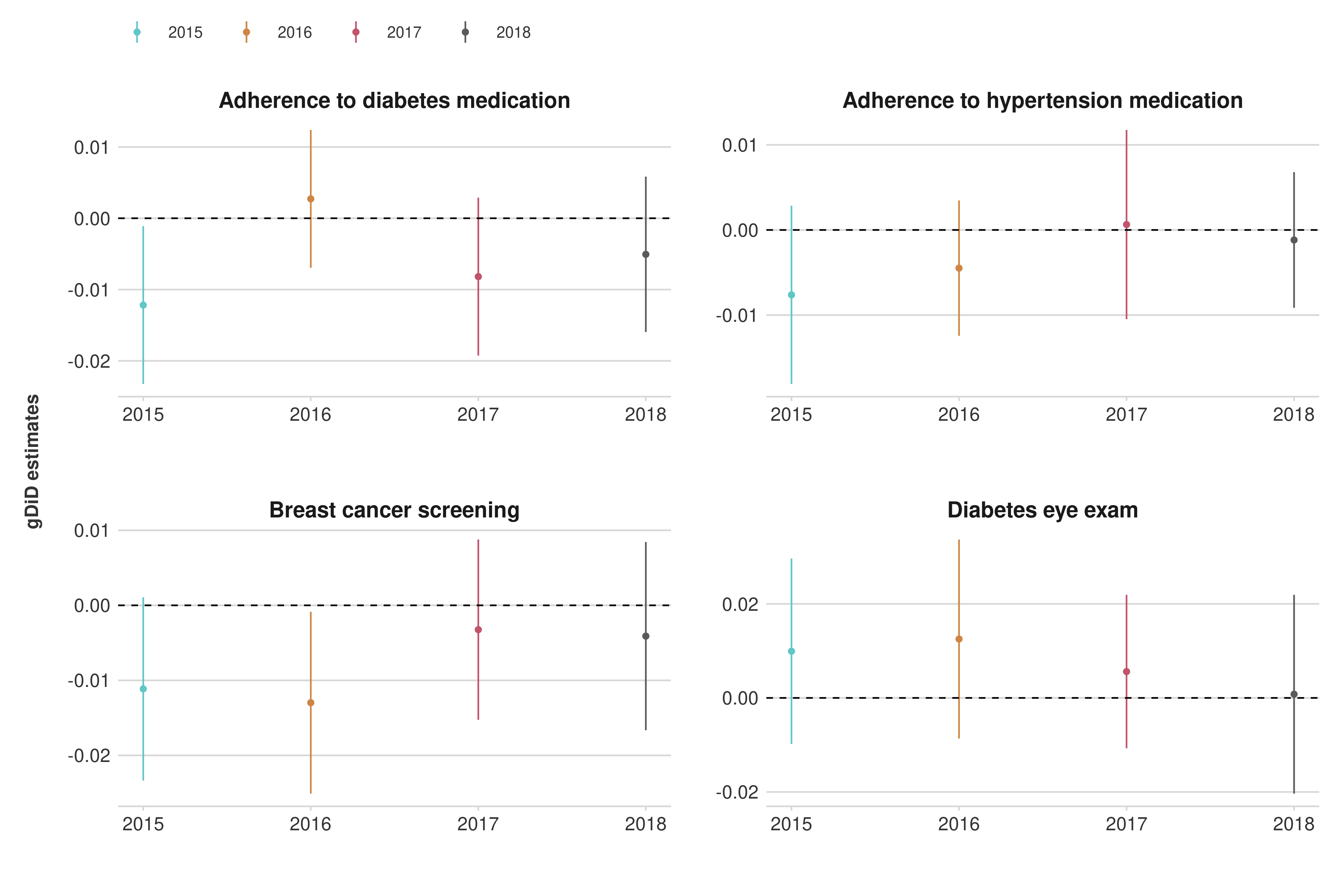}
\end{center}
\caption{Estimates of the effect of health system affiliation in the year of affiliation (2015-2018) on four health care quality measures. Points correspond to point estimates and line ranges correspond to 95\% multiplier bootstrap confidence intervals.}\label{real-data-fig}
\end{figure}

Aggregating across years, we estimate that affiliation has a sizable negative effect on breast cancer screening but smaller effects on the other outcomes. We estimate that the breast cancer screening rate is about 1.43 perecentage points lower in the year of affiliation than it would have been in the absence of affiliation (95\% CI [-2.43, -0.42]). We find that the naive ATT for affiliation (based on ignorability) in the year of affiliation is 3.71 percentage points (95\% CI [2.82, 4.61]), but in the year prior to affiliation, the ATT was even higher: 5.14 percentage points (95\% CI [3.88, 6.40]). For the other outcomes, we found smaller effects of -0.73 percentage points on adherence to diabetes medication, -0.06 percentage points on adherence to hypertension medication, and 0.36 percentage points on receipt of an eye exam among diabetes patients. More detailed results broken out by year of first affiliating with a health system are given in Figure \ref{real-data-fig}. Effects on breast cancer screening appear to have been stronger for POs who affiliated in 2015 and 2016.

\section{Discussion}\label{discussion}

We have proposed a very general approach for estimating treatment effects in longitudinal/panel data settings. Generalized DiD uses a stable bias assumption that is significantly weaker than either the no unmeasured confounding assumption underpinning ignorability estimators and significantly more flexible than the parallel trends assumption underpinning DiD estimators. Our approach is fully non-parametric and is implemented in an R package available at \texttt{https://github.com/mrubinst757/gdid}.

Generalized DiD also builds upon the ``in-time'' placebo tests frequently used in the synthetic controls literature. Specifically, \cite{abadie2015comparative} propose using pre-treatment data to predict pre-treatment outcomes as a type of validity check on the primary synthetic controls estimate. These tests produce error estimates that are essentially equivalent to the bias estimates that we propose. The key difference is that these tests are not used to correct the synthetic controls estimate -- instead, practitioners simply examine the pre-treatment errors to heuristically evaluate the validity of the synthetic controls estimate. There are at least two problems with this procedure generalized DiD solves: first, the pre-treatment errors may be large, suggesting that the primary estimate is invalid. Second, reporting the primary estimate conditional on small pre-treatment errors may lead to inflated Type I error rates, akin to issues identified with testing parallel trends in DiD \cite{roth2022pretest}. Generalized DiD solves this problem by essentially building the placebo tests into the estimation procedure, specifically by subtracting off some function of these ``placebo'' errors from the primary estimate. One can therefore think of generalized DiD as a more principled version of these ``in-time'' placebo tests.

On the other hand, the price of the stable bias assumption is a stronger positivity assumption (Assumption \ref{asmp:positivity}), which must be carefully considered when applying these methods. In settings where the positivity may be nearly violated, one may remove $\bYbar_{i0}$ from $\bW_{i0}$ and only condition on $\bX_i$ (as in the gDiD-0 estimator in Section \ref{sims}), or one may modify $\bW_{i0}$ to include functions of the pre-period outcomes for which there is better overlap, like the trends in the pre-period outcomes $Y_{it} - Y_{it-1}$ rather than the outcomes themselves (often done in the synthetic control literature\cite{abadie2021using}).

The connection between our estimator and typical ignorability estimators opens up a number of potentially interesting avenues for future research into potential violations of gDiD's required assumptions. Another potential way of addressing positivity violations is to use targeted maximum likelihood \cite{van2006targeted, schuler2017targeted}, which has been shown previously to handle positivity violations well \cite{porter2011relative}, instead of our approach based on augmented inverse probability weighting. Also, violations of the stable bias assumption could potentially be evaluated by adapting methods for omitted variable bias of causal effects under ignorability \cite{chernozhukov2022long}.

\bibliographystyle{plain}
\bibliography{did-bib}

\appendix

\section{Proof of Proposition 1 (main result)} \label{app:prop-1}
The proof proceeds by showing that the proposed estimator falls into the class of \textit{double machine learning} estimators of \cite{chernozhukov2018double}. First, we require a set of regularity conditions on the data-generating process. Define positive constants $C^*, c, \epsilon > 0$ and $\iota \in (0, 1)$.

\begin{assumption} We have the following\label{asmp:dgp-reg} for $t = 0,1$:
    \begin{enumerate}
        \item $\left(E|Y_{it}|^q\right)^{1/q} \leq C^*$.
        \item $\left\{E|Y_{it} - \mu_t(\bW{_it-1})|^2\right\}^{1/2} > c$.
        \item $\max E\left[\{Y_{it} - \mu_t(\bW_{it-1})\}^2 | \bW_{it-1}\right] \leq C$.
        \item $n\inv \sumin A_i \longrightarrow \iota$.
    \end{enumerate}
\end{assumption}

And we require a few further conditions on the nuisance function estimation
\begin{assumption}\label{asmp:nuisance-reg}
     For nuisance functions $\eta(\bW_{i0}) = \{\mu_1(\bW_{i0}), \mu_0(\bW_{i-1}), \pi(\bW_{i0}), \pi_0(\bW_{i-1})\}$  and estimated versions $\etahat(\bW_{i0}) = \{\muhat_1(\bW_{i0}), \muhat_0(\bW_{i-1}), \pihat(\bW_{i0}), \pihat_0(\bW_{i-1})\}$, we have $\left\{E|\etahat(\bW_{i0}) - \eta(\bW_{i0})|^q\right\}^{1/q} < C^*, \left\{E|\etahat(\bW_{i0}) - \eta(\bW_{i0})|^2\right\}^{1/2} = o_p(1),$ and $\max E\{\pihat(\bW_{i0}) - 1/2\} \leq 1/2 - \epsilon, \max E\{\pihat_0(\bW_{i-1}) - 1/2\} \leq 1/2 - \epsilon$.
\end{assumption}

Under these conditions, we have that the two components of $\tauhat$ are estimators of the form of \cite{chernozhukov2018double}, which implies that $\tauhat$ also falls into the same class of estimators:
\begin{align*}
    \tauhat &= \tauhat^{\text{ign.}}_1 - \tauhat^{\text{ign.}}_0\\
    \tauhat^{\text{ign.}}_1 &= n_1\inv\sumin \phi^{\text{ign.}}_1(\bO_i, \tau^{\text{ign.}}, \etahat)+ \tau^{\text{ign.}} = n_1\inv\sumin \left[Y_{i1}A_i - \frac{(1-A_i)\pihat(\bW_{i0})Y_{i1} + \{A_i - \pihat(\bW_{i0})\}\muhat_1(\bW_{i0})}{1-\pihat(\bW_{i0})}\right] \\
    \tauhat^{\text{ign.}}_0 &= n_1\inv\sumin \phi^{\text{ign.}}_0(\bO_i, \tau^{\text{ign.}}_0, \etahat) + \tau^{\text{ign.}}_0 = n_1\inv\sumin\left[Y_{i0}A_i +  \frac{(1-A_i)\pihat_0(\bW_{i-1})Y_{i0} + \{A_i - \pihat_0(\bW_{i-1})\}\muhat_0(\bW_{i-1})}{1-\pihat_0(\bW_{i-1})}\right]
\end{align*}
where $\tau^{\text{ign.}} = E(Y_{i1} |  A_i = 1) -  E\left\{E(Y_{i1} | \bW_{i0}, A_i = 0) | A_i = 1 \right\}$, and $\tau^{\text{ign.}}_0 = E(Y_{i0} |  A_i = 1) -  E\left\{E(Y_{i0} | \bW_{i-1}, A_i = 0) | A_i = 1 \right\}$.
These estimators satisfy Neyman orthogonality (see Definition 2.1 in \cite{chernozhukov2018double}) such that
\begin{align*}
    &E\left\{ \phi^{\text{ign.}}_1(\bO,\tau^{\text{ign.}}, \eta) \right\} = 0\\
    &E\left\{ \phi^{\text{ign.}}_0(\bO,\tau^{\text{ign.}}_0, \eta) \right\} = 0\\
    &\partial_\etatilde E\left\{ \phi^{\text{ign.}}_1(\bO,\tau^{\text{ign.}}, \eta) \right\} [\etatilde - \eta] = 0\\
    &\partial_\etatilde E\left\{ \phi^{\text{ign.}}_0(\bO,\tau^{\text{ign.}}_0, \eta) \right\} [\etatilde - \eta] = 0 \qquad \text{for all } \etatilde \in \Tsc_{\etatilde}
\end{align*}
where $\Tsc_{\etatilde}$ is a properly shrinking neighborhood of the true nuisance functions $\eta$ and $\partial_\eta E\left\{ \phi^{\text{ign.}}_1(\bO,\tau^{\text{ign.}}, \eta) \right\} [\etatilde - \eta]$ is the limit of the pathwise derivative
\begin{align*}
    \partial_\etatilde E\left\{ \phi(\bO,\tau,\eta) \right\} [\etatilde - \eta] &= \lim_{r \rightarrow0} \partial_r \left[ E\left\{ \phi(\bO,\tau,\eta + r(\etatilde - \eta)) \right\}\right].
\end{align*}

Thus, because $\tauhat = \tauhat^{\text{ign.}}_1 - \tauhat^{\text{ign.}}_0$ and the influence function for $\tau$, defined in \eqref{phi} is the difference between these two $\phi(\bO) = \phi(\bO, \tau, \eta) = \phi^{\text{ign.}}_1(\bO,\tau^{\text{ign.}}, \eta) - \phi^{\text{ign.}}_0(\bO,\tau^{\text{ign.}}_0, \eta)$, we obtain Neyman orthogonality for the gDiD estimator
\begin{align*}
    &E\left\{ \phi(\bO,\tau, \eta) \right\} = 0\\
    &\partial_\eta E\left\{ \phi(\bO,\tau, \eta) \right\} [\etatilde - \eta] = 0.
\end{align*}
This, along with the regularity conditions in Assumptions \ref{asmp:dgp-reg} and \ref{asmp:nuisance-reg}, ensures the results in the main text.

\section{Multiplier bootstrap variance estimation}\label{app:bootstrap}

In the text, we give plug-in estimators for the variance of $\tauhat$. In this section, we review the use of multiplier bootstrap estimators for the variance. 

\subsection{Standard iid setting}
The multiplier bootstrap proceeds by generating $n$ independent random variables $\bG = (G_i)_{i=1, ..., n}$ such that $E(G_i) = 1$ and $\var(G_i) = 1$. Then one bootstrap replicate can be computed as
\begin{align*}
    \tauhat^*_\bG = n_1\inv\sumin G_i \phihat(\bO_i).
\end{align*}
It is easy to see that $E(\tauhat^*_\bG | \bO) = n_1\inv\sumin \phihat(\bO_i) = \tauhat$, and $\var(\tauhat^*_\bG | \bO) = n_1^{-2}\sumin \phihat(\bO_i)^2$. So, $\var\{\nhalf(\tauhat_\bG^* - \tauhat)\}$ converges to $\sigma^2$ conditional on $\bO$. Therefore, the multiplier bootstrap may be used as an alternative to the plug-in estimator \eqref{sigmahat}. Operationally, this means generating a large number of replications $B$ of $\bG: \{\bG_b\}_{b = 1, .., B}$, yielding a large set of estimates $\{\tauhat^*_{\bG_b}\}_{1, ..., B}$. One may then construct a confidence interval for $\tau$ based on the quantiles of $\{\tauhat^*_{\bG_b}\}_{1, ..., B}$ or by estimating $\sigma^2$ as $\var\{\tauhat^*_{\bG_b}\}_{1, ..., B}$.

\subsection{Clustered data setting}

The multiplier bootstrap in the clustered data setting is very similar as in the iid setting, except that we now construct $C$ independent random variables $\bG = (G_j)_{j = 1, ..., C}$ and construct the bootstrap estimators as
\begin{align*}
    \tauhat_\bG = C\inv\sum_{j=1}^C G_j \phitilde(\bO_j).
\end{align*}
It is again easy to see that $E(\tauhat^*_\bG | \bO) = C\inv\sumin \phihat(\bO_j) = \tauhat$, and $\var(\tauhat^*_\bG | \bO) = C^{-2}\sum_{j=1}^C\phitilde(\bO_j)^2$. It is again straightforward to show that $C^{-1/2}(\tauhat^*_\bG - \tauhat) | \bO \longrightarrow N(0, \zeta^2).$

\section{Parametric nuisance function estimation}\label{app:parametric}

As noted in the text, in some cases we may wish to estimate all nuisance functions using parametric models (e.g. generalized linear models) and avoid data splitting altogether (e.g., if sample size is too small to feasibly allow sample splitting). In this case the asymptotic variance of the estimator depends on whether the models are correctly specified or not. However, regardless of whether the models were correctly specified, the asymptotic variance has a simple form assuming that the nuisance parameters and estimator were the solutions to estimating equations taking the form:
\begin{align*}
\mathbb{P}_n\begin{bmatrix}
m(\bO; \bbetahat) \\
\phi(\bO, \tauhat, \bbetahat),  
\end{bmatrix} = \mathbb{P}_n\left\{\varphi(\bO; \bthetahat)\right\} = 0
\end{align*}
where $\mathbb{P}_n\{f(\bO)\} = n\inv\sumin f(\bO_i)$, $\bbetahat$ is a finite-dimensional parameter, $m(\bO, \bbeta)$ is the estimating equation for the nuisance functions, $\varphi = (m, \phi)$, and $\bthetahat = (\tauhat, \bbetahat)$.  
For example, in the two time-period case, $m$ may be an estimating function, such as a score equation, for a series of models for $\eta = (\mu_1, \mu_0, \pi, \pi_0)$ parameterized by the finite-dimensional vector $\bbeta$. And $\phi$ is the estimated efficient influence function \eqref{phi} that plugs in nuisance function estimates parameterized by $\bbetahat$. Then, by Theorem 5.41 in van der Vaart (\cite{van2000asymptotic}):
\begin{align*}
\bthetahat - \btheta &= \begin{bmatrix}
\bbetahat \\
\tauhat
\end{bmatrix} - \begin{bmatrix}
\bbeta \\
\tau
\end{bmatrix} \\
&= \mathbb{P}_n\left\{-\left(\frac{\partial\mathbb{E}[\varphi(\bO; \btheta)]}{\partial \btheta}\right)^{-1}\varphi(\bO; \btheta) \right\} + o_p(1 / \sqrt{n})
\end{align*}
and, therefore, the asymptotic variance of $\bthetahat$ is given by:
\begin{align*}
\left(\frac{\partial\mathbb{E}[\varphi(\bO; \btheta)]}{\partial \btheta}\right)^{-1}\mathbb{E}[\varphi(\bO; \btheta) \varphi(\bO; \btheta) ^\top]\left(\frac{\partial\mathbb{E}[\varphi(\bO; \btheta)]}{\partial \btheta}\right)^{-1}
\end{align*}
The variance can be estimated by plugging-in the empirical counterparts (i.e. the sandwich estimator), and the variance estimate of $\hat{\tau}$ is given by the final diagonal element of this estimated matrix. We refer to \cite{van2000asymptotic} for more technical details.

\section{Proof of Proposition 2 (clustered treatment assignment)}\label{app:clustered}

The primary difference between the proof of Proposition 2 and Proposition 1 is the presence of clustering. We thus first establish that the number of clusters grows with sample size:
\begin{assumption}
    $C \rightarrow \infty$. 
\end{assumption}
Next, we restate the identifiability conditions (Assumptions \ref{asmp:consistency}, \ref{asmp:anticipation}, \ref{asmp:positivity}, and \ref{stable-bias}) in the clustered data setting.
\begin{assumption}[Consistency, clustered data setting]\label{asmp:consistency-cl}
    $Y_{ji1}(a) = Y_{ji1} \text{ when } A_j = a, a = 0, 1.$
\end{assumption}
\begin{assumption}[No anticipation, clustered data setting]\label{asmp:anticipation-cl}
    $Y_{jit}(a) = Y_{jit}, a = 0, 1; t < 1.$
\end{assumption}
\begin{assumption}[Positivity, clustered data setting]\label{asmp:positivity-cl}
    $P\left\{ p(\bWtilde_{j0}) < 1 - \epsilon\right\} = 1$ for some $\epsilon > 0$.
\end{assumption}
\begin{assumption}[Stable bias, clustered data setting]\label{stable-bias-cl}
    \begin{align*}E\left[E\left\{Y_{ji1}(0) | \bV_{ji0}, A_i = 1\right\} - E\left\{Y_{ji1} | \bV_{ji0}, A_i = 0 \right\} | A_i = 1\right] = E\left[E\left\{Y_{ji0} | \bV_{ji-1}, A_i = 1\right\} - E\left\{Y_{ji0} | \bV_{ji-1}, A_i = 0\right\} | A_i = 1\right].\end{align*}
\end{assumption}
These are mere extensions to the clustered data setting and do not introduce any notable additional complexity. 
Next, define cluster-level versions of the outcomes and mean functions
\begin{align*}
    Z_{jt} = \frac{C}{N_1}\sum_{i=1}^{n_j}Y_{jit}\\
    \mbar_{jt} = \frac{C}{N_1}\sum_{i=1}^{n_j}m_t(\bV_{jit-1})
\end{align*}
where $N_1 = \sum_{j=1}^C n_jA_j$ is the total number of treatment observations. Then we can rewrite $\tauhat$ in terms of these aggregate values
\begin{align*}
    \tauhat &= C\inv\sum_{j=1}^C \left[Z_{j1}A_j - \frac{(1-A_j)\phat(\bWtilde_{j0})Z_{j1} + \{A_j - \phat(\bWtilde_{j0})\}\widehat{\mbar}_1}{1-\phat(\bWtilde_{j0})} - \right.\\
    &\qquad\qquad \left.Z_{j0}A_j +  \frac{(1-A_j)\phat_0(\bWtilde_{j-1})Z_{j0} + \{A_j - \phat_0(\bWtilde_{j-1})\}\widehat{\mbar}_0}{1-\phat_0(\bWtilde_{j-1})}\right]\\
    &= C\inv\sum_{j=1}^C \phitilde(\bO_j).
\end{align*}
We may now analyze $\tauhat$ in the same manner as for Proposition 1 in Appendix \ref{app:prop-1}. We first establish the primitive regularity conditions
\begin{assumption} We have the following\label{asmp:dgp-reg-cl} for $t = 0,1$:
    \begin{enumerate}
        \item $\left(E|Z_{jt}|^q\right)^{1/q} \leq C^*$.
        \item $\left\{E|Z_{jt} - \mbar_t|^2\right\}^{1/2} > c$.
        \item $\max E\left[\{Z_{it} - \mbar_t\}^2 | \bV_{jit-1}\right] \leq C$.
        \item $C\inv \sumin A_j \longrightarrow \iota$.
    \end{enumerate}
\end{assumption}
And we adapt conditions on the nuisance function estimation
\begin{assumption}[Nuisance function convergence, clustered data setting]\label{asymptotic-convergence-assump-cl}
\begin{align*}
    &E\left[ \left\{\widehat{\mbar}_1 - \mbar_1 \right\}^2 \right]^{1/2} \times E\left[ \left\{\phat(\bWtilde_{j0}) - \pi(\bWtilde_{j0} \right\}^2 \right]^{1/2} = o_p(C^{-1/2})\\
    &E\left[ \left\{\widehat{\mbar}_0 - \mbar_0 \right\}^2 \right]^{1/2} \times E\left[ \left\{\pihat_0(\bW_{i-1}) - \pi_0(\bW_{i-1} \right\}^2 \right]^{1/2} = o_p(C^{-1/2})
\end{align*}
\end{assumption}
\begin{assumption}\label{asmp:nuisance-reg-cl}
     For nuisance functions $\eta(\bV_{ji0}) = \{\mbar_1, \mbar_0, p(\bWtilde_{j0}), p_0(\bWtilde_{j-1})\}$  and estimated versions $\etahat(\bV_{ji0}) = \{\widehat{\mbar}_1, \widehat{\mbar}_0, \phat(\bWtilde_{j0}), \phat_0(\bWtilde_{j-1})\}$, we have 
     \begin{itemize}
         \item $\left\{E|\etahat(\bV_{ji0}) - \eta(\bV_{ji0})|^q\right\}^{1/q} < C^*$
         \item $\left\{E|\etahat(\bV_{ji0}) - \eta(\bV_{ji0})|^2\right\}^{1/2} = o_p(1)$
         \item $\max E\{\phat(\bWtilde_{j0}) - 1/2\} \leq 1/2 - \epsilon, \max E\{\phat_0(\bWtilde_{j-1}) - 1/2\} \leq 1/2 - \epsilon$
     \end{itemize}, 
\end{assumption}

The argument proceeds just as in Appendix \ref{app:prop-1}.

\section{Complete simulation results}\label{app:sim-results}

\subsection{Simulation one}\label{app:sim-1}

We present additional results when setting $\zeta = 0.3, 0.5$. In these settings all ignorability-based estimators and gDiD perform poorly due to limited overlap between the pre-treatment outcome levels in the treatment and control groups.

\begin{table}\caption{Simulation study one results: $\zeta = 0.3$}
\centering
\begin{tabular}[t]{lllllllll}
\toprule
\multicolumn{1}{c}{ } & \multicolumn{4}{c}{Linear} & \multicolumn{4}{c}{Nonlinear} \\
\cmidrule(l{3pt}r{3pt}){2-5} \cmidrule(l{3pt}r{3pt}){6-9}
Estimator & Bias & RMSE & CIL & Coverage & Bias & RMSE & CIL & Coverage\\
\midrule
cDiD & -0.176 & 0.273 & 0.784 & 86.2 & -0.228 & 0.329 & 0.889 & 83\\
gDiD-0 & -0.179 & 0.272 & 0.791 & 85.7 & -0.218 & 0.316 & 0.872 & 83.3\\
DiD & -0.197 & 0.304 & 0.963 & 88.7 & -0.209 & 0.319 & 0.964 & 86.7\\
gDiD-2 & 0.4 & 3.095 & 2.354 & 46.3 & 0.542 & 1.063 & 1.494 & 39.5\\
gDiD-1 & 0.406 & 0.57 & 1.233 & 67 & 0.39 & 0.502 & 1.003 & 59.2\\
Ign-1 & -4.616 & 4.622 & 1.09 & 0 & -4.629 & 4.632 & 0.997 & 0\\
Ign-2 & -7.245 & 7.71 & 2.373 & 0.7 & -7.134 & 7.166 & 1.752 & 0.3\\
AugSynth & -9.854 & 9.896 & - & - & -9.772 & 9.819 & - & -\\
\bottomrule
\end{tabular}
\end{table}

\begin{table}\caption{Simulation study one results: $\zeta = 0.5$}
\centering
\begin{tabular}[t]{lllllllll}
\toprule
\multicolumn{1}{c}{ } & \multicolumn{4}{c}{Linear} & \multicolumn{4}{c}{Nonlinear} \\
\cmidrule(l{3pt}r{3pt}){2-5} \cmidrule(l{3pt}r{3pt}){6-9}
Estimator & Bias & RMSE & CIL & Coverage & Bias & RMSE & CIL & Coverage\\
\midrule
cDiD & -0.189 & 0.28 & 0.786 & 84.2 & -0.231 & 0.337 & 0.896 & 80.3\\
gDiD-0 & -0.19 & 0.281 & 0.792 & 84.2 & -0.226 & 0.324 & 0.874 & 78.7\\
DiD & -0.211 & 0.329 & 0.965 & 84.7 & -0.214 & 0.328 & 0.964 & 85.7\\
gDiD-1 & 0.632 & 0.911 & 1.48 & 50 & 0.606 & 0.833 & 1.359 & 46.2\\
Ign-1 & -7.644 & 7.653 & 1.661 & 0 & -7.686 & 7.696 & 1.639 & 0\\
AugSynth & -16.301 & 16.328 & - & - & -16.26 & 16.29 & - & -\\
Ign-2 & -11310059838.405 & 800877093417.37 & 427364180154.257 & 3 & 303.724 & 6421.427 & 1264.803 & 2.2\\
gDiD-2 & -38559885534.194 & 2106347444114.77 & 1105410193452.54 & 39 & -70686491808.493 & 2145878599062.18 & 1308869531810.81 & 34.9\\
\bottomrule
\end{tabular}
\end{table}

\subsection{Simulation two}\label{app:sim-2}

The table below presents all simulation results from all specifications explored.

\begin{landscape}
\begin{longtable}{lllllllllllll}
\toprule
\multicolumn{1}{c}{ } & \multicolumn{4}{c}{Linear} & \multicolumn{4}{c}{Additive} & \multicolumn{4}{c}{Nonlinear} \\
\cmidrule(l{3pt}r{3pt}){2-5} \cmidrule(l{3pt}r{3pt}){6-9} \cmidrule(l{3pt}r{3pt}){10-13}
Estimator & Bias & RMSE & CIL & Coverage & Bias & RMSE & CIL & Coverage & Bias & RMSE & CIL & Coverage\\
\midrule
\endfirsthead
\multicolumn{13}{@{}l}{\textit{(continued)}}\\
\toprule
Estimator & Bias & RMSE & CIL & Coverage & Bias & RMSE & CIL & Coverage & Bias & RMSE & CIL & Coverage\\
\midrule
\endhead

\endfoot
\bottomrule
\endlastfoot
\addlinespace[0.3em]
\multicolumn{13}{l}{\textbf{$\gamma = 0$: $\beta_t = \beta$}}\\
\hspace{1em}AugSynth & 0.24 & 0.255 & 0.963 & 100 & - & - & - & - & 0.228 & 0.241 & 0.911 & 100\\
\hspace{1em}DID & -0.001 & 0.087 & 0.351 & 94.8 & - & - & - & - & 0.019 & 0.093 & 0.357 & 94.6\\
\hspace{1em}Ign-1 & 0.456 & 0.464 & 0.349 & 0 & - & - & - & - & 0.401 & 0.409 & 0.328 & 0\\
\hspace{1em}Ign-2 & 0.315 & 0.327 & 0.344 & 5.8 & - & - & - & - & 0.285 & 0.297 & 0.321 & 6.4\\
\hspace{1em}gDID-1 & 0.011 & 0.13 & 0.496 & 92.8 & - & - & - & - & 0.023 & 0.126 & 0.489 & 94.8\\
\hspace{1em}gDID-2 & 0.005 & 0.126 & 0.486 & 94.6 & - & - & - & - & 0.023 & 0.121 & 0.467 & 94.8\\
\hspace{1em}gDiD-0 & -0.001 & 0.105 & 0.401 & 94 & - & - & - & - & 0.017 & 0.101 & 0.337 & 90.8\\
\hspace{1em}sDID & -0.002 & 0.159 & 0.605 & 94.2 & - & - & - & - & 0.02 & 0.157 & 0.618 & 94.6\\
\addlinespace[0.3em]
\multicolumn{13}{l}{\textbf{$\gamma = 0$: $\beta_t - \beta_{t-1} = d$}}\\
\hspace{1em}AugSynth & 0.425 & 0.439 & 1.701 & 100 & - & - & - & - & 0.413 & 0.424 & 1.651 & 100\\
\hspace{1em}DID & 0.422 & 0.432 & 0.368 & 0.8 & - & - & - & - & 0.387 & 0.398 & 0.375 & 1\\
\hspace{1em}Ign-1 & 0.593 & 0.603 & 0.445 & 0 & - & - & - & - & 0.493 & 0.503 & 0.406 & 0\\
\hspace{1em}Ign-2 & 0.46 & 0.473 & 0.425 & 1.4 & - & - & - & - & 0.385 & 0.398 & 0.388 & 3.4\\
\hspace{1em}gDID-1 & -0.066 & 0.168 & 0.616 & 93.4 & - & - & - & - & -0.083 & 0.169 & 0.593 & 93.8\\
\hspace{1em}gDID-2 & -0.132 & 0.197 & 0.577 & 83.6 & - & - & - & - & -0.133 & 0.193 & 0.545 & 81.6\\
\hspace{1em}gDiD-0 & 0.244 & 0.266 & 0.418 & 38 & - & - & - & - & 0.188 & 0.212 & 0.344 & 45\\
\hspace{1em}sDID & 0.019 & 0.148 & 0.606 & 96 & - & - & - & - & 0.001 & 0.156 & 0.618 & 96\\
\addlinespace[0.3em]
\multicolumn{13}{l}{\textbf{$\gamma_t = \gamma$: $\beta_t = \beta$}}\\
\hspace{1em}AugSynth & 0.273 & 0.285 & 1.091 & 100 & 0.122 & 0.169 & 0.558 & 100 & 0.49 & 0.562 & 1.982 & 100\\
\hspace{1em}DID & 0.833 & 0.837 & 0.336 & 0 & 0.103 & 0.148 & 0.427 & 84 & 0.398 & 0.422 & 0.547 & 17.4\\
\hspace{1em}Ign-1 & 0.287 & 0.3 & 0.345 & 10.8 & 0.368 & 0.378 & 0.372 & 1.8 & 0.132 & 0.733 & 1.093 & 59.8\\
\hspace{1em}Ign-2 & 0.281 & 0.296 & 0.344 & 12.2 & 0.337 & 0.35 & 0.366 & 4.8 & -1.766 & 40.604 & 8.424 & 66\\
\hspace{1em}gDID-1 & -0.096 & 0.158 & 0.467 & 84.4 & -0.034 & 0.134 & 0.528 & 96 & -0.131 & 0.709 & 1.066 & 95\\
\hspace{1em}gDID-2 & -0.081 & 0.157 & 0.483 & 89 & -0.055 & 0.138 & 0.497 & 93.2 & -1.969 & 40.105 & 8.262 & 94.4\\
\hspace{1em}gDiD-0 & 0.39 & 0.4 & 0.347 & 1.2 & 0.218 & 0.242 & 0.376 & 38.4 & 0.344 & 0.507 & 0.684 & 32.4\\
\hspace{1em}sDID & 0.01 & 0.096 & 0.351 & 93.4 & -0.122 & 0.205 & 0.647 & 88 & 0.038 & 0.169 & 0.66 & 95\\
\addlinespace[0.3em]
\multicolumn{13}{l}{\textbf{$\gamma_t = \gamma$: $\beta_t - \beta_{t-1} = d$}}\\
\hspace{1em}AugSynth & 0.434 & 0.448 & 1.736 & 100 & 0.437 & 0.456 & 1.75 & 100 & 0.609 & 0.692 & 2.452 & 100\\
\hspace{1em}DID & 1.625 & 1.63 & 0.512 & 0 & 0.623 & 0.634 & 0.474 & 0 & 0.716 & 0.733 & 0.638 & 0.4\\
\hspace{1em}Ign-1 & 0.561 & 0.572 & 0.446 & 0.6 & 0.631 & 0.641 & 0.474 & 0.2 & 0.088 & 4.598 & 2.06 & 36.6\\
\hspace{1em}Ign-2 & 0.455 & 0.468 & 0.427 & 2.2 & 0.572 & 0.585 & 0.464 & 0.4 & 0.163 & 1.591 & 1.624 & 46.6\\
\hspace{1em}gDID-1 & -0.202 & 0.242 & 0.507 & 64.4 & -0.098 & 0.178 & 0.589 & 89 & -0.314 & 4.506 & 1.999 & 96.2\\
\hspace{1em}gDID-2 & -0.19 & 0.24 & 0.58 & 76.4 & -0.077 & 0.168 & 0.577 & 91 & -0.224 & 1.562 & 1.527 & 95.6\\
\hspace{1em}gDiD-0 & 1.01 & 1.018 & 0.489 & 0 & 0.573 & 0.584 & 0.414 & 0.4 & 0.682 & 0.736 & 0.652 & 7.1\\
\hspace{1em}sDID & 0.396 & 0.406 & 0.366 & 1.2 & -0.081 & 0.179 & 0.656 & 93 & 0.195 & 0.252 & 0.688 & 83.4\\
\addlinespace[0.3em]
\multicolumn{13}{l}{\textbf{$\gamma_t - \gamma_{t-1} = c$: $\beta_t = \beta$}}\\
\hspace{1em}AugSynth & 0.252 & 0.265 & 1.008 & 100 & 0.05 & 0.145 & 0.464 & 100 & 0.522 & 0.614 & 2.116 & 100\\
\hspace{1em}DID & 1.38 & 1.385 & 0.468 & 0 & 0.283 & 0.307 & 0.451 & 32.6 & 0.654 & 0.677 & 0.697 & 3.2\\
\hspace{1em}Ign-1 & 0.253 & 0.268 & 0.342 & 18.8 & 0.333 & 0.348 & 0.391 & 7.6 & 0.027 & 2.392 & 1.52 & 71.2\\
\hspace{1em}Ign-2 & 0.255 & 0.27 & 0.343 & 17 & 0.303 & 0.319 & 0.385 & 12 & 0.027 & 1.793 & 1.433 & 76.8\\
\hspace{1em}gDID-1 & -0.072 & 0.141 & 0.474 & 89.6 & -0.02 & 0.145 & 0.539 & 93.8 & -0.236 & 2.366 & 1.491 & 93.8\\
\hspace{1em}gDID-2 & -0.069 & 0.14 & 0.476 & 89.8 & -0.033 & 0.139 & 0.505 & 94.2 & -0.177 & 1.77 & 1.347 & 93\\
\hspace{1em}gDiD-0 & 0.414 & 0.425 & 0.378 & 1.2 & 0.192 & 0.227 & 0.405 & 52.6 & 0.447 & 0.536 & 0.735 & 35.4\\
\hspace{1em}sDID & 0.57 & 0.577 & 0.359 & 0 & -0.113 & 0.201 & 0.649 & 90 & 0.231 & 0.291 & 0.695 & 74.8\\
\addlinespace[0.3em]
\multicolumn{13}{l}{\textbf{$\gamma_t - \gamma_{t-1} = c$: $\beta_t - \beta_{t-1} = d$}}\\
\hspace{1em}AugSynth & 0.44 & 0.452 & 1.759 & 100 & 0.381 & 0.41 & 1.525 & 100 & 0.584 & 0.696 & 2.37 & 100\\
\hspace{1em}DID & 2.276 & 2.282 & 0.669 & 0 & 0.771 & 0.781 & 0.51 & 0 & 0.957 & 0.979 & 0.8 & 0.2\\
\hspace{1em}Ign-1 & 0.47 & 0.483 & 0.434 & 2.4 & 0.527 & 0.539 & 0.488 & 0.8 & 0.292 & 0.563 & 1.065 & 52.4\\
\hspace{1em}Ign-2 & 0.437 & 0.452 & 0.43 & 3.4 & 0.498 & 0.511 & 0.477 & 0.6 & 0.231 & 0.442 & 1.113 & 66.4\\
\hspace{1em}gDID-1 & -0.181 & 0.23 & 0.535 & 72.4 & -0.089 & 0.177 & 0.591 & 91.4 & -0.108 & 0.472 & 1.045 & 96\\
\hspace{1em}gDID-2 & -0.192 & 0.246 & 0.575 & 74.2 & -0.072 & 0.174 & 0.569 & 90.4 & -0.149 & 0.387 & 1.027 & 95\\
\hspace{1em}gDiD-0 & 1.029 & 1.039 & 0.544 & 0 & 0.5 & 0.517 & 0.441 & 2 & 0.737 & 0.805 & 0.771 & 12.3\\
\hspace{1em}sDID & 1.016 & 1.022 & 0.423 & 0 & -0.071 & 0.183 & 0.659 & 94.2 & 0.375 & 0.412 & 0.745 & 48.8\\
\addlinespace[0.3em]
\multicolumn{13}{l}{\textbf{$\gamma_t - \gamma_{t-1} = c_t$: $\beta_t - \beta_{t-1} = d_t$}}\\
\hspace{1em}AugSynth & 0.529 & 0.538 & 2.114 & 100 & 0.57 & 0.578 & 2.279 & 100 & 0.421 & 0.439 & 1.686 & 100\\
\hspace{1em}DID & 0.793 & 0.807 & 0.574 & 0.2 & 0.475 & 0.492 & 0.492 & 3.4 & 0.078 & 0.174 & 0.629 & 92.2\\
\hspace{1em}Ign-1 & 1.168 & 1.173 & 0.468 & 0 & 0.849 & 0.855 & 0.434 & 0 & 0.457 & 0.47 & 0.428 & 0.8\\
\hspace{1em}Ign-2 & 0.54 & 0.55 & 0.395 & 0.2 & 0.535 & 0.543 & 0.391 & 0 & 0.21 & 0.242 & 0.398 & 39.2\\
\hspace{1em}gDID-1 & 0.629 & 0.64 & 0.504 & 0 & 0.262 & 0.296 & 0.537 & 52.4 & -0.061 & 0.173 & 0.63 & 95\\
\hspace{1em}gDID-2 & 0.001 & 0.136 & 0.511 & 93.2 & -0.03 & 0.131 & 0.522 & 95.4 & -0.271 & 0.327 & 0.644 & 62.8\\
\hspace{1em}gDiD-0 & 0.007 & 0.13 & 0.502 & 93 & 0.149 & 0.184 & 0.391 & 65.4 & -0.212 & 0.259 & 0.515 & 61.6\\
\hspace{1em}sDID & -0.557 & 0.585 & 0.694 & 11.6 & -0.846 & 0.864 & 0.738 & 0.4 & -0.702 & 0.733 & 0.872 & 12\\*
\end{longtable}
\end{landscape}

\end{document}

%% file: GrandMacros.tex
\def\bzero{{\bf 0}}

\def\ba{{\mbox{\boldmath$a$}}}

\def\bv{{\bf v}}
\def\bw{{\bf w}}
\def\bx{{\bf x}}
\def\by{{\bf y}}

\def\bA{{\bf A}}

\def\bG{{\bf G}}

\def\bO{{\bf O}}

\def\bV{{\bf V}}
\def\bW{{\bf W}}
\def\bX{{\bf X}}
\def\bY{{\bf Y}}
\def\bZ{{\bf Z}}

\def\thick#1{\hbox{\rlap{$#1$}\kern0.25pt\rlap{$#1$}\kern0.25pt$#1$}}

\def\bbeta{\boldsymbol{\beta}}

\def\btheta{\boldsymbol{\theta}}

\def\bomega{\boldsymbol{\omega}}

\def\smbalpha{\boldsymbol{{\scriptstyle{\alpha}}}}

\def\mhat{{\widehat m}}

\def\phat{{\widehat p}}

\def\bWtilde{{\widetilde \bW}}

\def\etahat{{\widehat\eta}}

\def\muhat{{\widehat\mu}}

\def\pihat{{\widehat\pi}}

\def\sigmahat{{\widehat\sigma}}
\def\tauhat{{\widehat\tau}}

\def\phihat{{\widehat\phi}}

\def\etatilde{{\widetilde\eta}}

\def\phitilde{{\widetilde\phi}}

\def\bbetahat{{\widehat\bbeta}}

\def\bthetahat{{\widehat\btheta}}

\def\smbalpha{\widehat{\smbalpha}}

\def\abar{\bar{ a}}

\def\hbar{\bar{ h}}

\def\mbar{\bar{ m}}

\def\Abar{\bar{ A}}

\def\babar{\bar{ \ba}}

\def\bybar{\bar{ \by}}

\def\bAbar{\bar{ \bA}}

\def\bWbar{\bar{ \bW}}

\def\bYbar{\bar{ \bY}}

\def\Asc{{\cal A}}

\def\Psc{{\cal P}}

\def\Tsc{{\cal T}}

\def\half{\frac{1}{2}}

\def\nhalf{n^{\half}}

\def\nnhalf{n^{-\half}}

\def\var{\mbox{var}}

\def\sumin{\sum_{i=1}^n}

\def\inv{^{-1}}

\def\var{\mbox{var}}

\def\mybox#1{\vskip1mm \begin{center}
        \hspace{.0\textwidth}\vbox{\hrule\hbox{\vrule\kern6pt
\parbox{.9\textwidth}{\kern6pt#1\vskip6pt}\kern6pt\vrule}\hrule}
        \end{center} \vskip-5mm}
\def\lboxit#1{\vbox{\hrule\hbox{\vrule\kern6pt
      \vbox{\kern6pt#1\vskip6pt}\kern6pt\vrule}\hrule}}

\def\thickboxit#1{\vbox{{\hrule height 1mm}\hbox{{\vrule width 1mm}\kern6pt
          \vbox{\kern6pt#1\kern6pt}\kern6pt{\vrule width 1mm}}
               {\hrule height 1mm}}}

\def\fat#1{\hbox{\rlap{$#1$}\kern0.25pt\rlap{$#1$}\kern0.25pt$#1$}}